\documentclass[11pt]{article}
\pdfoutput=1

\usepackage{euscript}
\usepackage{amssymb}
\usepackage{amsfonts}
\usepackage{amsbsy}
\usepackage{epsfig}
\usepackage{amsthm}
\usepackage{amscd}
\usepackage{amstext}
\usepackage{verbatim}
\usepackage{amsmath}
\usepackage{cancel}
\usepackage{capt-of}
\usepackage{empheq}
\usepackage{subfigure}
\usepackage{xcolor}

\usepackage[nosort]{cite}
\usepackage{bm}
\usepackage{authblk}
\usepackage{esint}
\usepackage[T1]{fontenc}

\usepackage[pdftex,linktoc=all]{hyperref}
\hypersetup{ 
colorlinks=true, 
linkcolor=black, 
citecolor=red, 
}

\def\ben{\begin{equation}}
\def\een{\end{equation}}
\def\half{{\textstyle{\frac{1}{2}}}}
    
  \let\q=\theta 
\let\l=\lambda

\let\C=\Chi

\let\pa=\partial
\def\be{\begin{equation}}
\def\ee{\end{equation}}
\def\beq{\begin{equation}}
\def\eeq{\end{equation}}
\def\ba{\begin{array}}
\def\ea{\end{array}}

\def\dalemb#1#2{{\vbox{\hrule height .#2pt
       \hbox{\vrule width.#2pt height#1pt \kern#1pt
               \vrule width.#2pt}
       \hrule height.#2pt}}}

\newcommand{\bea}{\begin{eqnarray}}
\newcommand{\eea}{\end{eqnarray}}
\DeclareMathOperator{\tr}{tr}
\DeclareMathOperator{\sech}{sech}
\newcommand{\Tr}{{\rm Tr} }

\makeatletter
\newcommand*\bigcdot{\mathpalette\bigcdot@{.5}}
\newcommand*\bigcdot@[2]{\mathbin{\vcenter{\hbox{\scalebox{#2}{$\m@th#1\bullet$}}}}}
\makeatother

\renewcommand{\eqref}[1]{(\ref{#1})}

\def\R{{{\mathbb R}}}
\def\C{{{\Bbb C}}}

\title{Entanglement in the Quantum Hall Matrix Model}
\author{Alexander Frenkel$^\sharp$ and Sean A. Hartnoll$^{\flat}$}
\affil{\it $^\sharp$Department of Physics, Stanford University, \\
\it Stanford, CA 94305-4060, USA \\
\it $^\flat$Department of Applied Mathematics and Theoretical Physics, \\
\it University of Cambridge, Cambridge CB3 0WA, UK
}
\date{}

\begin{document}

\maketitle

\begin{abstract}

Characterizing the entanglement of matrix degrees of freedom is essential for understanding the holographic emergence of spacetime. The Quantum Hall Matrix Model is a gauged $U(N)$ matrix quantum mechanics with two matrices whose ground state is known exactly and describes an emergent spatial disk with incompressible bulk dynamics. We define and compute an entanglement entropy in the ground state associated to a cut through the disk. There are two contributions. A collective field describing the eigenvalues of one of the matrices gives a gauge-invariant chiral boundary mode leading to an expected logarithmic entanglement entropy. Further, the cut through the bulk splits certain `off-diagonal' matrix elements that must be duplicated and associated to both sides of the cut. Sewing these duplicated modes together in a gauge-invariant way leads to a bulk `area law' contribution to the entanglement entropy. All of these entropies are regularized by finite $N$.

\end{abstract}

\newpage

\tableofcontents

\newpage

\section{Emergent geometry, entanglement and  matrices}
\label{sec:intro}

In holographic and string theoretic approaches to gravity, spacetime emerges from a many-body quantum mechanical system. The `pre-geometric' degrees of freedom are typically large matrices of oscillators or extended strings. While a great deal has been learnt about matrix and string theories, it remains to be understood how the quantum state of the microscopic degrees of freedom allows the emergence of gravity as a collective phenomenon.

An important clue into the microscopic nature of quantum states that support semiclassical gravity comes from the recent understanding of the entanglement entropy in such states \cite{Ryu:2006bv, Hubeny:2007xt, Faulkner:2013ana, Engelhardt:2014gca}. In particular, there must be a large, universal entanglement associated to spatial partitions of the emergent semiclassical spacetime.

The difficulty in making use of the entanglement clue is that the microscopic degrees of freedom are not spatially local. Matrices can be thought of as `fuzzy' particles while strings are inherently extensive. It is unclear, then, what factorization of the microscopic Hilbert space corresponds to a local partition of the emergent space. In this work we will study this question within a relatively simple gauged $U(N)$ matrix quantum mechanics with two matrices \cite{Polychronakos:2001mi}, whose ground state is known explicitly \cite{Hellerman:2001rj}.

The very simplest matrix quantum mechanics involve a single matrix \cite{Klebanov:1991qa}. A single matrix can be diagonalized, leading to a many-particle problem for the eigenvalues. Consequently, the singlet-sector entanglement can be captured using conventional methods \cite{Das:1995vj, Das:1995jw, Hartnoll:2015fca}. Going beyond a single matrix, however, the degrees of freedom cannot all be reduced to eigenvalues and one must understand how to partition the `off-diagonal' modes of the matrices.

The starting point of our work is the recent observation \cite{Das:2020jhy, Das:2020xoa, Hampapura:2020hfg} that it is natural to diagonalize one of the multiple matrices. The matrix quantum mechanics that are best known to lead to emergent spacetimes, such as the BFSS and BMN models \cite{Banks:1996vh, Berenstein:2002jq}, already contain a hint of the emergent space: The matrices $X_1, X_2, \ldots$ in these models are in correspondence with the coordinates $x_1, x_2, \ldots$ in the emergent space. Therefore a partition of the emergent space that can be written as $f(\vec x) > 0$ can be pulled back (up to an ordering prescription) to the matrix variables as $F \equiv f(\vec X) > 0$. This partition can be implemented as a partition on the eigenvalues $\left\{F_a\right\}_{a=1}^N$ of $F$. The simplest case is a partition along a coordinate plane $x > x_o$. This corresponds to partitioning the eigenvalues of $X$ according to whether they are less or greater than $x_o$, much as was done in the models with a single matrix. The entanglement in the eigenvalues again reduces to a conventional many-particle problem.
However, we are still left with the conceptual problem of what to do with the remaining degrees of freedom, as these cannot be reduced to eigenvalues.

In \cite{Das:2020jhy, Das:2020xoa, Hampapura:2020hfg} various scenarios for partitioning the remaining modes were considered. We will take our cue from the observation in \cite{Hampapura:2020hfg} that the `off-diagonal' (i.e.~non-eigenvalue) modes should be treated using methods developed to address closely related issues of factorization of the Hilbert space in lattice gauge theory. We review these issues in the following paragraph. In this work we will go on to consider a specific multi-matrix quantum mechanics where the state of the off-diagonal modes is fully determined by a Gauss law. This gives us a simple, explicit setup allowing us to focus on the conceptual challenges.

The basic gauge-invariant variables of conventional lattice gauge theories --- Wilson loops ---  are spatially extended objects. Wilson loops that extend across the geometric entanglement cut cannot be associated purely to one side of the cut. It was proposed in \cite{Donnelly:2011hn}, see also the closely related discussion in \cite{Buividovich:2008gq, Casini:2013rba, Ghosh:2015iwa}, that the degrees of freedom living on a link that extends across the cut should be duplicated and that one copy be
assigned to each side of the cut. Prior to obtaining the reduced quantum state describing one of the sides, the quantum state of the full system must be uplifted into an extended Hilbert space with the duplicated link. Requiring this uplifted state to be gauge invariant forces the two copies to be highly entangled. This then leads to a gauge-theoretic contribution to the entanglement entropy of the cut. In simple situations the duplicated entanglement cut modes are maximally entangled so that the entropy $s_\text{cut} \sim \ell \, \log (\dim R)$, with $\ell$ the number of cut links and $\dim R$ the dimension of the Hilbert space on each link.

In our matrix quantum mechanics, the analogous objects to degrees of freedom living on a link extending across the cut are `off-diagonal' matrix elements \cite{Hampapura:2020hfg}. These connect eigenvalues on opposites sides of the partition of eigenvalues that we have described above. As with the lattice gauge theory, we will therefore proceed to duplicate these degrees of freedom, and embed the state in an extended Hilbert space. The precise form of the duplication will be determined by how a partition of the underlying classical phase space necessarily leads to new `boundary cut' modes, in the spirit of \cite{Donnelly:2016auv}. We will again find that gauge invariance forces a large entanglement between the two factors of the extended Hilbert space.

\begin{figure}[h!]
    \centering
    \includegraphics[width=0.55\textwidth]{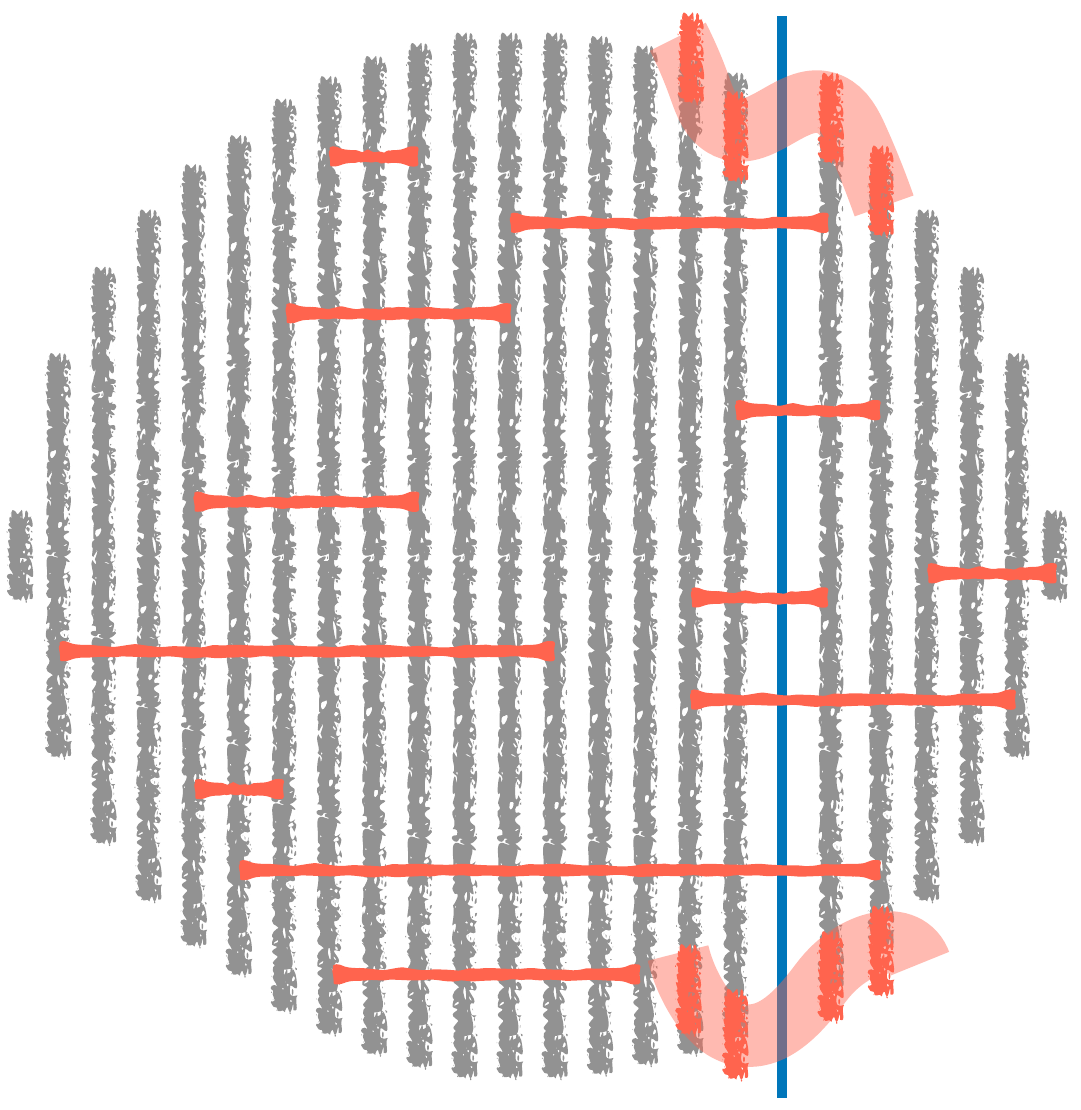}
    \caption{The eigenvalues of the $X$ matrix are distributed in a circle and delocalized in the vertical direction. These are shown as fuzzy gray lines in the figure. The vertical entanglement cut through the droplet is shown with a solid blue line. Several illustrative off-diagonal modes connecting pairs of eigenvalues are shown with horizontal red lines. If these cross the entanglement cut they will contribute to the entanglement. The collective dynamics of the eigenvalues themselves is incompressible and described by a boundary mode. Fluctuations of this boundary mode close to the cut, shown in red in the figure, also contribute to the entanglement.}
    \label{fig:circleplot}
\end{figure}

Figure \ref{fig:circleplot} illustrates how the two types of entanglement --- of eigenvalues and of off-diagonal modes --- work out in the Quantum Hall Matrix Model. We will introduce the model properly in the following section. The model has two $N \times N$ matrices $X$ and $Y$ that are canonically conjugate to each other. The geometric cut is built on a partition of the eigenvalues of $X$ and therefore corresponds to a vertical cut through the emergent spatial droplet in this model (we will also discuss other cut geometries shortly). These eigenvalues are delocalized in the vertical $Y$ direction. We will show that the quantum state of the eigenvalues is that of a chiral boson on the boundary of the droplet. We will thereby obtain a logarithmic entanglement between the eigenvalues.
The remaining degrees of freedom in this model are unitary matrices that can be thought of as parametrizing the off-diagonal modes of $Y$. These are pure gauge modes of the full system and their ground state is trivial. Nonetheless, defining a geometric partition forces some of these modes to be duplicated and then entangled. We will show that, after carefully accounting for permutation symmetries, this leads to a counting problem for the gauge-theoretic contribution to the entanglement entropy that is solved by the Hardy-Ramanujan formula. This gauge-theoretic entanglement is thereby found to be proportional to the length of the cut through the droplet. This `area law' entanglement is consistent with an emergent gauge-theoretic locality in the bulk of the droplet. Both the boundary and bulk contributions to the entanglement are regulated by the finite $N$ of the underlying matrices.

In units where the emergent droplet has unit radius and the length of the entanglement cut through the droplet is $L$, the full entanglement entropy we find will be
\be\label{eq:answer}
s = \frac{(Nk)^{1/2} \log (N L)}{\sqrt{6}} L + \frac{1}{6} \log \left( N L \right) + \cdots \,.
\ee
These are the leading terms from the bulk (gauge theoretic) and boundary (collective field) entanglement, respectively. 
In (\ref{eq:answer}) the integer $k$ is the charge carried by the ground state under a certain $U(1) \subset U(N)$ and we have dropped non-universal order one constants that appear inside the logarithms. As we discuss in section \ref{sec:gauge}, it is possible that subleading bulk terms are larger than the boundary term. The entanglement in (\ref{eq:answer}) is consistent with the well-known fact that this model exhibits many features of the quantum Hall effect \cite{Polychronakos:2001mi, Tong:2015xaa}. The second, logarithmic, term is precisely the entanglement of the expected boundary chiral boson with a short distance regulator determined by $N$, cf.~\cite{Belin:2019mlt}. The first term is a bulk `area law' (with a multiplicative logarithmic violation, that we discuss shortly) that might be expected from an emergent Chern-Simons field in the bulk of the droplet, cf.~\cite{PhysRevLett.96.110404,Das:2015oha,Wong:2017pdm}. However, the microscopic matrix dynamics that regulates the Chern-Simons field has no manifest bulk locality, and so this behavior is a priori nontrivial.

In conventional realizations of the quantum Hall effect from a microscopically local lattice model, the emergence of a Chern-Simons field is captured by the topological entanglement entropy \cite{PhysRevLett.96.110404, PhysRevLett.96.110405}. This is a constant correction to the area law term. We have not been able to compute the subleading bulk term reliably in (\ref{eq:answer}) due to technical complications with the vertical cut that are described later. To address this point, we have also considered a circular cut. That is, we cut out a circle of circumference $C_o$ from the interior of the droplet. This 
cut does not intersect the boundary of the droplet and
offers some technical simplifications (at large $k$), allowing us to calculate the subleading correction to the gauge-theoretic entanglement. We obtain
\begin{align}\label{eq:answer2}
s^\text{circular}_\text{cut}  = \frac{(Nk)^{1/2}}{\sqrt{6}} C_o - \log (\sqrt{3} N k \, C_o^2/\pi^2)  + \cdots \,. 
\end{align}
There are two comments to make here. Firstly, there is no multiplicative logarithmic violation of the leading order area law term in this case. We will demonstrate later that the logarithmic violation of the area law in (\ref{eq:answer}) is due to the fact that the vertical entanglement cut intersects the boundary of the droplet. Secondly, the additive logarithmic correction in (\ref{eq:answer2}) is gauge-theoretic in origin, unlike the collective field logarithm in (\ref{eq:answer}). It has some similarities to the topological entanglement entropy, but is not obviously connected to the conventional $- \half \log k$ term of abelian Chern-Simons theory. It is not clear that any topological terms emerging from the matrix quantum mechanics should match those of local lattice models, and we will discuss this term further in section \ref{sec:topological}.

Our main results (\ref{eq:answer}) and (\ref{eq:answer2}) demonstrate the emergence of area law entanglement in matrix quantum mechanics. In section \ref{sec:moyal} we argue that the area law holds for any cut geometry: the entanglement is rooted in the gauge-theoretic nonlocality of the quantum state, which is robustly encoded in a dominant singular value of the off-diagonal matrix degrees of freedom. While there is no emergent dynamical gravity in the simple model that we have solved, indeed there are no propagating bulk excitations at all, the entanglement that we have uncovered may be a precursor of the universal gravitational entanglement reflected in the Ryu-Takayanagi formula \cite{Ryu:2006bv, Hubeny:2007xt, Faulkner:2013ana, Engelhardt:2014gca}.

\section{The model and the ground state}
\label{sec:model}

The effective long-wavelength description of an  incompressible quantum Hall fluid \cite{PhysRevLett.50.1395} is naturally expressed in terms of area-preserving diffeomorphisms. This leads to the idea that a minimal microscopic realization of quantum Hall physics may be obtained by discretizing these diffeomorphisms into a $U(N)$ symmetry at large but finite $N$ \cite{Susskind:2001fb}.

To obtain a model with $N$ finite, it is necessary to add an external potential that constrains the fluid to a finite droplet in the plane \cite{Polychronakos:2001mi}. For nice discussions of the physics of this model see e.g. \cite{Polychronakos:2001mi, Tong:2015xaa}. We will refer to this theory as the Quantum Hall Matrix Model. The Hamiltonian is simply given by a confining potential term
\be\label{eq:H}
H = Z^\dagger_{ab} Z^{\phantom{\dagger}}_{ba} \,.
\ee
The indices $a,b$ run from $1$ to $N$ and $Z_{ab} = \frac{1}{\sqrt{2}} \left(X_{ab} + i Y_{ab} \right)$,
where $X$ and $Y$ are Hermitian matrices whose components obey the quantum mechanical commutators $[X_{ab},Y_{cd}] = i \delta_{ad} \delta_{bc}$. Physical states are in addition required to carry a specific $U(N)$ charge, such that they are annihilated by
\be\label{eq:gauss}
G_{ac} \equiv - i \left(X_{ab} Y_{bc} -  X_{bc} Y_{ab} \right) + \Psi^{\phantom{\dagger}}_a \Psi^{\dagger}_{c} - k \, \delta_{ac} \,.
\ee
Here $k=1,2,3, \ldots$ and the $\Psi_a$ form a complex vector with quantum mechanical commutator $[\Psi^{\phantom{\dagger}}_{\vphantom{b}a},\Psi^\dagger_b] = \delta_{ab}$. We have operator ordered such that the trace of the constraint is $ \Psi^{\phantom{\dagger}}_a \Psi^{\dagger}_{a} = N k$, consistent with the conventions in e.g.~\cite{Tong:2015xaa}.
This specifies a nontrivial $U(1) \subset U(N)$ charge of the state, carried by the $\Psi$ modes. While these modes do not have any nontrivial dynamics --- and will not play a significant role in this paper --- the background charge that they supply supports the ground state. The traceless part of the constraint requires that physical states are $SU(N)$ singlets.

The exact ground state of this model was found in \cite{Hellerman:2001rj}. In our conventions it is given by 
\be
|\psi\rangle = \left[\epsilon^{a_1 \ldots a_N} \Psi^\dagger_{a_1}
(\Psi^\dagger Z^\dagger)_{a_2} \cdots (\Psi^\dagger Z^{\dagger \, N-1})_{a_N}
\right]^{k-1} |0\rangle \,.
\ee
Here $|0\rangle$ is annihilated by $Z_{ab}$ and $\Psi_a$. We saw above that $X$ and $Y$ are canonically conjugate. Therefore the wavefunction may be represented as a function of the coordinate $X$ and coherent state label $\phi$, where $\Psi_a |\phi\rangle = \phi_a |\phi\rangle$. Following the discussion in the introductory section \ref{sec:intro} above, we would like to furthermore express $X$ in terms of a unitary matrix $U$ and eigenvalues $x$. Thus we write
\be\label{eq:diag}
X_{ab} = U^{\phantom{\dagger}}_{\vphantom{b}ac} x^{\phantom{\dagger}}_{\vphantom{b}c} U^\dagger_{cb} \,, \qquad \phi_a = U_{ab} \widetilde \phi_{b} \,. 
\ee
In terms of these variables the ground state wavefunction becomes \cite{Karabali:2001xq}
\be\label{eq:gs}
\psi(U,\widetilde \phi,x) = \left( \det U \right)^{k-1} \, \prod_{a<b} (x_a - x_b)^{k-1} e^{-\frac{1}{2}\sum_d x_d^2} \, \prod_{c} \widetilde \phi_c^{k-1} e^{-\frac{1}{2} \sum_d |\widetilde \phi_d|^2} \,.
\ee
This wavefunction is to be normalized as
\be\label{eq:meas}
\int dU d^N\widetilde \phi d^Nx \prod_{a<b} (x_a - x_b)^2 |\psi|^2 = 1 \,,
\ee
where $dU$ is the Harr measure on $U(N)$ and the eigenvalue term is the usual Vandermonde determinant, which arises in the measure when the matrix is parametrized as in (\ref{eq:diag}).

The wavefunction (\ref{eq:gs}) has factorized into three pieces that may be discussed independently. The variables $x_a$ and $\widetilde \phi_a$ are gauge-invariant data while $U$ parametrizes gauge orbits. The $(\det U)^{k-1}$ term is a phase associated to the nontrivial $U(1)$ charge of the state. As part of the diagonalization of the matrix $X$ in (\ref{eq:diag}) we may order the eigenvalues $x_1 \leq x_2 \leq \cdots \leq x_N$. The possible minus signs arising from exchanging two eigenvalues in the term $(x_a - x_b)^{k-1}$ in (\ref{eq:gs}) are repackaged into the signs picked up by $(\det U)^{k-1}$ upon exchanging two columns of $U$.

There is one further piece of notation to introduce. A $U(1)^N$ subgroup of $U(N)$ is carried by the individual components of $\Psi$. This charge was removed upon rotating to the gauge-invariant $\widetilde \Psi$, but it will occasionally be useful to re-instate the phase degrees of freedom by setting
\be\label{eq:psihat}
\hat \Psi_a = e^{i \theta_a} \widetilde \Psi_a \,, \qquad i \theta_a \equiv (\log U)_{aa} \,.
\ee
In terms of these variables the overall determinant in (\ref{eq:gs}) is absorbed in the $\hat \Psi_a$ wavefunction:
\be\label{eq:gshat}
(\det U)^{k-1} \prod_c \widetilde \phi^{k-1}_c = e^{i (k-1) \sum_a \theta_a} \prod_c \widetilde \phi^{k-1}_c = \prod_c \hat \phi^{k-1}_c \,.
\ee

\section{The collective field and boundary mode entanglement}
\label{sec:col}

\subsection{Collective field and large $N$ saddle point}

We first study the eigenvalue part of the wavefunction (\ref{eq:gs}). This is precisely the ground state wavefunction of the Calogero model for $N$ particles moving in a quadratic potential in one dimension, see \cite{Polychronakos:2001mi, Karabali:2001xq} and references therein. The physics of this state will be most transparent in terms of the collective field \cite{JEVICKI1980511, ANDRIC1983307}
\be\label{eq:collective}
n(x) \equiv \sum_a \delta(x - x_a) \,.
\ee
The collective field is invariant under permutation of the eigenvalues. As we have already ordered the eigenvalues, this redundancy has been accounted for. The eigenvalue part of the wavefunction (\ref{eq:gs}) becomes, incorporating the Vandermonde term from the measure (\ref{eq:meas}) into the wavefunction to produce a shift $k-1 \to k$,
\be\label{eq:psin}
\psi[n] = e^{S[n]} \,,
\ee
where
\be\label{eq:S}
S[n] = \frac{k}{2} \int dx_1 dx_2 n(x_1) n(x_2) \log |x_1 - x_2| - \frac{1}{2} \int dx n(x) x^2  \,.
\ee
There are no phases in this wavefunction, again due to the ordering of eigenvalues in (\ref{eq:gs}). Thus upon symmetrizing the integrals over $x$ in (\ref{eq:S}) the argument of the logarithm has acquired an absolute value.

The wavefunction (\ref{eq:psin}) is to be normalized as
\be\label{eq:norm2}
\int {\mathcal D} n \, J[n] \psi[n]^2 = 1 \,.
\ee
The additional measure $J[n]$ arises due to the change of variables (\ref{eq:collective}) from the eigenvalues $\{x_a\}$ to the collective field $n(x)$. It is given by
\begin{align}
J[n] & = \int d^Nx \delta[n(z) - \sum_a \delta(z-x_a)] = \int d^Nx {\mathcal D} \lambda e^{i \int dx \lambda(x) [n(x) - \sum_a \delta(x-x_a)] } \nonumber \\
& = \int {\mathcal D} \lambda e^{i \int dx \lambda(x) n(x)} \int d^Nx  e^{- i \sum_a \lambda(x_a)} =  \int {\mathcal D} \lambda e^{i \int dx \lambda(x) n(x)} \left[ \int dx e^{- i \lambda(x)} \right]^N \nonumber \\
& = \int {\mathcal D} \lambda e^{i \int dx \lambda(x) n(x) + N \log \left[ \int dx \, e^{- i \lambda(x)} \right]} \,. \label{eq:J}
\end{align}
In the first line $z$ is just a dummy variable for the function $n$. The form obtained in the last line is well-suited to a saddle point evaluation at large $N$.

The large $N$ support of the wavefunction can be obtained by using the measure (\ref{eq:J}) in (\ref{eq:norm2}) and simultaneously performing the $n$ and $\lambda$ integrals in a saddle point expansion. This is done explicitly in Appendix \ref{sec:appA}. It is found that the only role of the measure $J[n]$ at leading orders in the large $N$ expansion is to impose normalization and positivity constraints: $\int dx n(x) = N$ and $n(x) \geq 0$.

With the normalization and positivity constraints, it is well-known that the functional (\ref{eq:S}) is extremized on the Wigner semi-circle distribution \cite{Brezin:1977sv}
\be\label{eq:wigner}
n_o(x) = \frac{2 N}{\pi R^2} \sqrt{R^2 - x^2} \,, \qquad R^2 = 2N k \,.
\ee
In the quantum Hall context, this distribution of $X$ eigenvalues is interpreted as arising from a uniform two dimensional circular droplet of radius $R$. The radius agrees with that in \cite{Polychronakos:2001mi}. As we noted in the discussion around Fig.~\ref{fig:circleplot}, the $X$ eigenvalues are delocalized in the vertical direction.

\subsection{Fluctuations and chiral boundary mode}

Fluctuations about the large $N$ saddle (\ref{eq:wigner}) are described by
\be
n(x) = n_o(x) + \delta n(x) \,.
\ee
The wavefunction for the fluctuations is obtained from (\ref{eq:psin}) and (\ref{eq:S}) to be
\be\label{eq:psi3}
\psi[\delta n] = e^{\frac{k}{2} \int dx_1 dx_2 \delta n(x_1) \delta n(x_2) \log |x_1 - x_2|} \,.
\ee
The divergence at $x_1 = x_2$ is integrable.
We show in Appendix \ref{sec:appA} that the measure term $J[n]$ in (\ref{eq:J}) imposes the normalization condition $\int dx \delta n(x) = 0$ and requires $\delta n$ to vanish outside the support $x \in [-R,R]$.

We will now see that the wavefunction (\ref{eq:psi3}) can be expressed in terms of a chiral edge mode on the boundary of the droplet. To this end we follow some manipulations in \cite{JACKIW1981133}. Firstly set $x = R \cos \theta$, with $\theta \in [0,\pi]$ and note that
\be
\sum_{m=1}^\infty \frac{\cos (m \q_1) \cos (m \q_2)}{m} =  - \frac{1}{2} \log |2 (\cos \q_1 - \cos \q_2)| \,.
\ee
Then the wavefunction becomes
\be\label{eq:cos}
\psi[\delta n] =  e^{ - \frac{1}{\pi}\sum_{m=1}^\infty \frac{1}{m} \left[ \int d\theta \varphi(\theta) \cos(m \theta) \right]^2} \,.
\ee
Here we defined for convenience
\be
\varphi(\theta) \equiv \sqrt{\pi k} R \sin \theta \, \delta n (R \cos \theta) \,.
\ee
Various constant terms in the exponent were removed using the normalization constraint $\int d\theta \varphi(\theta) = 0$. The wavefunction (\ref{eq:cos}) is not manifestly local in $\theta$. Locality in $\theta$ can be revealed in several steps, as follows.

Firstly, introduce an integral over a new field $\phi(\theta)$, so that
\begin{align}
\psi[\delta n] & = \int {\mathcal D} \phi \, e^{-  \frac{1}{\pi} \sum_m m [\int d\theta \phi(\theta) \cos(m \theta)]^2 - i \int d\theta \phi(\theta) \varphi(\theta)} \label{eq:n1a}\\
& = \int {\mathcal D} \phi \, \Phi[\phi] e^{- i \int d\theta \phi(\theta) \varphi(\theta) } \,. \label{eq:n1}
\end{align}
To go from (\ref{eq:n1a}) to (\ref{eq:cos}) use the identity $\pi \delta(\theta_1 - \theta_2) = 2 \sum_m \cos(m \theta_1) \cos(m \theta_2) + 1$ to introduce a sum into the final term in (\ref{eq:n1a}). The constant term is again removed using 
the constraint $\int d\theta \varphi(\theta) = 0$. Relatedly, the field $\phi$ that has been introduced has no constant mode (we will come back to this point later).\footnote{This $\phi$ is unrelated to the coherent state coordinates $\phi_a$ introduced in the previous section. These variables will never appear in the same discussion.} We have not kept track of the overall normalization of the wavefunction.
In the second line (\ref{eq:n1}) we recognized the first term in the exponent in (\ref{eq:n1a}) as the ground state functional of a local Hamiltonian
\be\label{eq:HH}
H_\text{bdy} \Phi[\phi] \equiv \left( \int_0^\pi d\theta  \left[\half \Pi(\theta)^2 + \half \phi'(\theta)^2 \right] \right) \Phi[\phi] = E_o \Phi[\phi] \,.
\ee
Here $[\phi(\theta),\Pi(\theta')] = i \left( \delta(\theta-\theta') - \frac{1}{\pi}\right)$. The factor of $\frac{1}{\pi}$ is due to the absence of a constant mode in the field, so that the integral over $\theta$ or $\theta'$ gives zero. To obtain (\ref{eq:n1a}) one can expand $\phi(\theta)$ and $\Pi(\theta)$ as a sum over cosines 
\be
\phi(\theta) = \sum_{m=1}^\infty \phi_m \cos(m \theta) \,, \qquad
\Pi(\theta) = \frac{2}{\pi} \sum_{m=1}^\infty \pi_m \cos(m \theta) \,.
\ee
Here $[\phi_m,\pi_n] = i \delta_{mn}$. The Hamiltonian in (\ref{eq:HH}) becomes
\be\label{eq:H3}
H_\text{bdy} = \frac{1}{\pi} \sum_{m=1}^\infty \left(\pi_m^2 + \frac{m^2 \pi^2}{4} \phi_m^2 \right) = \sum_{m=1}^\infty m a_m^\dagger a_m \,.
\ee
where we dropped the zero-point energy and defined $a_m = \sqrt{m \pi} (  \phi_m + \frac{i}{2 m \pi} \pi_m)$. These obey the usual $[a_m,a_n^\dagger] = \delta_{mn}$. The Harmonic oscillator ground state of these modes is given by $\Phi[\phi]$ in (\ref{eq:n1a}) and (\ref{eq:n1}).

The Hamiltonian in (\ref{eq:H3}) describes a left- and a right-moving mode on a semicircle. These can be re-interpreted as a single right-moving mode on the full circle (the choice of chirality here is arbitrary). This will be the chiral mode on the boundary of the quantum Hall droplet. Define the right moving field, cf. \cite{SONNENSCHEIN1988752},
\be
\phi_R(\theta) = \sum_{m=1}^\infty \frac{1}{2 \sqrt{m \pi}} \left(a_m e^{- i m \theta} + a_m^\dagger e^{i m \theta} \right) \,. \label{eq:phiR}
\ee
Where now $\theta \in [0,2\pi]$, so that
\be
H_\text{bdy} = \int_0^{2\pi} d\theta \left(\phi_R'(\theta) \right)^2 \,.
\ee
As shown in \cite{SONNENSCHEIN1988752}, this Hamiltonian corresponds to the Floreanini-Jackiw Lagrangian density for a chiral boson on a circle
\be\label{eq:FJ}
{\mathcal L}_\text{bdy} = - \dot \phi_R \phi'_R - (\phi_R')^2 \,.
\ee

We can use the Lagrangian (\ref{eq:FJ}) to write the ground state $\Phi$ as a Euclidean path integral over a half-cylinder $\R^+ \times S^1$, with an $S^1$ boundary at Euclidean time $\tau=0$. The original wavefunction (\ref{eq:n1}) finally becomes
\be\label{eq:n2}
\psi[\delta n] = \int {\mathcal D}\phi_R e^{- \int d\tau d\theta \left[i \pa_\tau \phi_R \pa_\theta \phi_R + (\pa_\theta \phi_R)^2 \right]- i \int d\theta \phi_R(\theta) \varphi_R(\theta) } \,.
\ee
Here the final integral is over the $S^1$ that is the boundary of the half-cylinder ($\tau=0$). This has coordinate $\theta \in [0,2\pi]$. We defined
$\varphi_R(\theta) = \frac{1}{2} \varphi(\theta)$ for $\theta \in [0,\pi]$ and $\varphi_R(\theta) = \frac{1}{2} \varphi(2\pi-\theta)$ for $\theta \in [\pi,2\pi]$. The final coupling in (\ref{eq:n2}) is obtained from the fact that
\be
\phi_R(\theta) + \phi_R(2\pi - \theta) = 2 \sum_m \phi_m \cos(m \theta) = 2 \phi(\theta) \,.
\ee
The collective field wavefunction (\ref{eq:n2}) is the (field space) `Fourier transform' of the ground state wavefunction of a chiral boson on a circle. This transform is local in the $\theta$ coordinate and therefore will not affect the entanglement associated to a partition of $\theta$ (up to a normalization constraint that we discuss shortly).

\subsection{Boundary entanglement}

We wish to obtain the entanglement entropy associated to a partition of the eigenvalues into $x < x_o$ and $x > x_o$. In the original matrix quantum mechanics this is a `target space entanglement' \cite{Mazenc:2019ety} of the eigenvalues. In Appendix \ref{sec:appB} we show how to map this entanglement onto a `base space entanglement' of the collective field $n(x)$. This correspondence has previously been discussed in \cite{Sugishita:2021vih}. A normalization constraint nonetheless remains so that the traces of the reduced density matrix, obtained by tracing over eigenvalues with $x<x_o$, are given by
\be\label{eq:trace3}
\tr \rho_\text{red}^n = \int \prod_{i=1}^n {\mathcal D} n_i \mu\left[n_i, \widetilde n_i \right] \overline \psi\left[n_i\right] \psi\left[\widetilde n_{i}\right] \,.
\ee
Here $n_i = n_i^< + n_i^>$ and $\widetilde n_{i} = n_i^< + n_{i+1}^>$, with $n_i^<$ the projection of $n_i$ to $x < x_o$ and $n_i^>$ the projection to $x > x_o$. That is, $\widetilde n_i$ mixes the replicas. We define $n_{n+1}^> = n_1^>$. The measure $\mu\left[n_i, \widetilde n_i \right]$ enforces that $n_i$ and $\widetilde n_i$ both integrate to $N$.

It is easily verified that replica-symmetric copies of the Wigner semicircle (\ref{eq:wigner}) give a large $N$ saddle point of (\ref{eq:trace3}) that is furthermore the absolute maximum of the exponent, see Appendix \ref{sec:appC}. It follows that at leading order $\tr \rho_\text{red}^n = (\tr \rho_\text{red})^n = 1$ once the reduced density matrix is normalized. Therefore, any entanglement is contained within the wavefunction (\ref{eq:n1}) for the fluctuations $\delta n_i$.
For convenience we can rescale the fluctuations $\delta n_i \to \delta n_i/\sqrt{\pi k}$, as this rescaling
cancels from the trace once the density matrix is normalized. Thus we can take
\be\label{eq:psi4}
\psi[\delta n_i] = \int {\mathcal D} \phi_i \Phi[\phi_i] e^{- i \int_{-R}^R dx \, \delta n_i(x) \phi_i(x)} \,.
\ee
In Appendix \ref{sec:appB} we use (\ref{eq:psi4}) to compute the traces (\ref{eq:trace3}). The only subtlety has to do with normalization constraints. It turns out that the Lagrange multipliers enforcing these constraints combine with the bosonic fields $\phi_i$ to re-instate the constant modes of these fields in each region. We will denote the fields with constant mode included $\phi^F$ (with $F$ standing for `full'). The upshot is that performing the integrals over the $\delta n_i$ one obtains
\be
\tr \rho_\text{red}^n = \sqrt{n} \int \prod_{i=1}^n \frac{{\mathcal D} \phi^F_i {\mathcal D} \widetilde \phi^F_i}{\text{vol}\left(\textstyle \frac{1}{\pi} \int d\theta \widetilde \phi^F_{n} \right)} \overline \Phi[\phi^F_i] \Phi[\widetilde \phi^F_i] \, \delta\left[ \phi_i^{F<} - \widetilde \phi_i^{F<} \right] \delta\left[\phi_i^{F>} -  \widetilde \phi_{i-1}^{F>} \right]  \,. \label{eq:trfinalmain}
\ee
This is precisely the entanglement of the bosonic state $\Phi$ under a spatial partition, as defined via the replica trick, with $\phi^F_i = \phi_i^{F<} + \phi_i^{F>}$ and $\widetilde \phi^F_i = \widetilde \phi_i^{F<} + \widetilde \phi_i^{F>}$. Note that the tildes now denote an independent field that is independently integrated over. The overall factor of the volume of the constant mode cancels out the contribution of this mode to the entanglement. It is natural to consider this constant mode to be compact --- because the field is massless and hence the wavefunction of the constant mode is non-normalizable if non-compact --- and the answer is independent of the field range used to regulate the expression. The overall factor of $\sqrt{n}$ in (\ref{eq:trfinalmain}) gives a subleading contribution to the entropies at large $N$ and will be dropped in the remainder.

From our discussion in the previous section, showing that $\Phi[\phi]$ is the ground state of a chiral boson on a circle, the traces (\ref{eq:trfinalmain}) will give the R\'enyi entropies
\be
s_n \equiv \frac{1}{1-n} \log \frac{\tr \rho_\text{red}^n}{\left(\tr \rho_\text{red} \right)^n} \,,
\ee
of a chiral boson CFT on a circle, with the constant mode excluded. Upon mapping to a circle, the region $x > x_o$ corresponds to the arc $- \theta_o < \theta < \theta_o$, with $x_o = R \cos \theta_o$. The computation of $s_n$ for this region falls within the classic CFT analyses of \cite{Holzhey:1994we, Lunin:2000yv}. The answer for the R\'enyi entropies is therefore
\be
s_n = \frac{1}{12} \left(1 + \frac{1}{n} \right) \log \left[ \frac{2 R}{\epsilon} \sin \theta_o \right] = \frac{1}{12} \left(1 + \frac{1}{n} \right) \log \left[ \frac{2 R}{\epsilon}\sqrt{1 - \left(\frac{x_o}{R}\right)^2} \right] \,. \label{eq:sn}
\ee
Here $\epsilon$ is a short distance cutoff, which we proceed to discuss in more detail.

Within the collective field approach, the short distance
cutoff $\epsilon$ is a non-perturbative property of the measure $J[n]$ in (\ref{eq:J}). This property is not easily incorporated in the large $N$ expansion that we have performed. Fortunately, it is known from other approaches to this system \cite{Polychronakos:2001mi, Tong:2015xaa} --- as well as in closely related matrix theories \cite{Berenstein:2004kk, Itzhaki:2004te} --- that the effect of finite $N$ is to truncate the mode expansion of the chiral field at $m=N$ in (\ref{eq:phiR}). We have seen the chiral field emerge from the collective excitations of the gauge-invariant eigenvalues. Alternatively, let $Z_\text{cl}$ be the classical matrix ground state. Perturbations of the ground state that obey the gauge constraint take the form $Z(t) = Z_\text{cl} + \sum_{m=1}^N c_m(t)\, (Z_\text{cl}^\dagger)^{m-1}$. Substituting into the Hamiltonian, the $c_m(t)$ are seen to describe a chiral boson with momentum truncated at $m=N$ \cite{Polychronakos:2001mi, Tong:2015xaa}. These two different descriptions must, of course, agree on the gauge-invariant physical modes of the system. Recall that we are using a collective field description because the target space partition of the eigenvalues at $x = x_o$ directly leads to a corresponding factorization of the collective field. In the following section we will furthermore see how this eigenvalue partition allows us to identify the gauge-theoretic edge modes associated with the extension of the partition into the bulk of the droplet. However, the alternate classical mode perspective is the easiest way to see the mode truncation.

With the truncation just described, the ratio $R/\epsilon$ in (\ref{eq:sn}) is of order $N$, uniform around the boundary circle, and does not depend on $k$. It follows that the finite boundary mode contribution to the Von Neumann entropy (the first R\'enyi entropy, $s \equiv s_1$) is
\be\label{eq:ent}
s_\text{bdy} = \frac{1}{6} \log \left( N L \right) + \cdots \,. 
\ee
Here
\be\label{eq:len}
L = 2 \sqrt{1 - \left(\frac{x_o}{R}\right)^2} \,.
\ee
is the length of the entanglement cut through the droplet, as in Fig.~\ref{fig:circleplot}, in units where the circle has radius one. The nonuniversal $\cdots$ terms are order one and do not have any singular dependence on $x_o/R$ or $N$. A closely analogous logarithmic entanglement from eigenvalues was found previously in \cite{Das:1995vj, Das:1995jw, Hartnoll:2015fca, Sugishita:2021vih}, using the fact that when $k=1$ the state can be described in terms of free fermions. The prefactor of the logarithm is $1/6$ in (\ref{eq:ent}) rather than $1/3$ because the boundary mode on the circle is chiral, in agreement with the result obtained from a computation in Chern-Simons theory \cite{Belin:2019mlt}. This result could also have been obtained from the wavefunction $\Phi[\phi]$ for left- and right-moving modes on a semicircle, showing that the (UV regulated) state we are considering does not suffer from obstructions to defining the entanglement of chiral theories \cite{Hellerman:2021fla}.

\section{The bulk entanglement}

\subsection{Preliminary comments}

With the transformation (\ref{eq:diag}) we have chosen to express the Hilbert space of the theory in terms of functions of the eigenvalues $x$, unitary matrices $U$ and vectors $\widetilde \Psi$. The ground state (\ref{eq:gs}) factorizes between these three variables. In the previous section we have shown that the ground state of the eigenvalues $x$ can be mapped onto the wavefunction (\ref{eq:n2}) for a chiral boson on a circle and that a `target space' partition of the eigenvalues captures the geometric entanglement of this chiral mode according to (\ref{eq:ent}). The entanglement is regulated by the graininess of the underlying eigenvalue description, which truncates the Fourier modes of the chiral boson. In this section we turn to the entanglement contained in the unitary matrices. The vectors $\widetilde \Psi$ will be seen not to contribute to the entanglement.

The unitary matrices are pure gauge modes of the full system and their wavefunction is trivial (recall from (\ref{eq:gshat}) that the overall determinant of $U$ in the wavefunction can be absorbed into the $\hat \Psi$). If one gauge-fixes prior to quantization, i.e. restricting to the space of gauge orbits, there will be no gauge-theoretic contribution to the entanglement.
Such a restriction would, however, limit the kinds of partitions of the system that are possible. It is useful to keep in mind the case of conventional gauge theories with extended Wilson loop observables. A geometric partition of the bulk cannot be described by simply partitioning the gauge-invariant observables in that case. As we recalled in section \ref{sec:intro} above, given a cut of the spatial geometry, Wilson lines that cross the cut must be duplicated, associated to both regions and entangled in a gauge-invariant way. In cases where the bulk degrees of freedom are pure gauge (as in e.g.~Chern-Simons theory) then these duplicated modes correspond to a subset of the gauge transformations that act nontrivially on the geometric boundary. These are the `boundary cut' modes.

Given the close connection of our matrix model to the quantum Hall effect \cite{Polychronakos:2001mi, Tong:2015xaa}, it is natural to suspect that the unitary matrices are related to the pure gauge degrees of freedom in the interior of the quantum Hall droplet (that are effectively described by Chern-Simons theory).
We would like, then, to identify the `boundary cut' modes that arise in the matrix model upon making a geometric cut across the bulk of the droplet. This bulk cut extends the cut that we have already defined --- by partitioning the eigenvalues of $X$ --- on the boundary circle. For reasons that we explain later in this section, we propose that the boundary cut modes are given by
\be\label{eq:Ubdy}
U = U_L U_R \in \frac{U(N-M)}{S_{N-M}} \times \frac{U(M)}{S_M} \subset U(N)\,.
\ee
Here $M$ is the number of $X$ eigenvalues on the right side of the cut. The transformations (\ref{eq:Ubdy}) act on left and right sub-blocks of the matrices, with permutations quotiented out in order to retain the ordering of the eigenvalues of $X$ in each region.
The remainder of this section will have two parts. Firstly, we will motivate the choice of modes in (\ref{eq:Ubdy}). Secondly, we will show that, upon quantization, the entanglement of these modes is given by a counting problem that we will solve. The volume of the phase space of the modes (\ref{eq:Ubdy}) will depend upon the classical large $N$ eigenvalue distribution discussed in the previous section, that we can think of as having built the space. As we summarized in section \ref{sec:intro} this counting problem leads to a bulk area law entanglement, consistent with an emergent two dimensional locality in the quantum mechanical state.

\subsection{Matrix block partition}
\label{sec:block}

The boundary partition of the eigenvalues corresponds to assigning each eigenvalue to $x_L$ or $x_R$ depending on whether it is less or greater than $x_o$. This partition of the eigenvalue space then induces a factorization on the space of functions of eigenvalues, which is the Hilbert space of the boundary modes. As we discussed in the introduction, and in the spirit of \cite{Das:2020jhy, Das:2020xoa, Hampapura:2020hfg}, the eigenvalue partition can be `uplifted' to a block partition of matrices. This will, in turn, induce a factorization on the Hilbert space of functions of matrices. The first step is to write the $X$ matrix very explicitly as
\be\label{eq:X}
X =  U \left(
\begin{array}{cc}
    x_L & 0 \\
    0 & x_R
\end{array}
\right)
U^\dagger \,,
\ee
Here $x_L$ and $x_R$ are the partitioned, ordered eigenvalues. Any Hermitian $X$ can be parametrized in this way. It will be important that the eigenvalues are ordered in this parametrization.

We can similarly partition the matrix $Y$ by setting
\be
Y = U Y^\text{cl} U^\dagger \,.
\ee
Imposing the Gauss law constraint (\ref{eq:gauss}) with $X$ diagonalized as in (\ref{eq:X}) fixes \cite{Polychronakos:2001mi}
\be\label{eq:yclass}
Y^\text{cl}_{ab} = y_a \delta_{ab}  - i \frac{\widetilde \Psi_b \widetilde \Psi^\dagger_a}{x_a - x_b} \,.
\ee
The diagonal components $y_a$ are unfixed while the second term is only present for $a \neq b$. By imposing Gauss's law at the level of the matrices themselves we are restricting to the physical phase space already at a classical level. We will proceed to quantize this space shortly.
The corresponding block partition is then
\be\label{eq:Y}
Y =  U \left(
\begin{array}{cc}
    Y^\text{cl}_{LL} & Y^\text{cl}_{LR} \\
    Y^\text{cl}_{RL} & Y^\text{cl}_{RR}
\end{array}
\right)
U^\dagger \,.
\ee
Note that $Y^\text{cl}_{LL} = Y^{\text{cl}\,\dagger}_{LL}$ only depends on $x_L$, $Y^\text{cl}_{RR}$ only depends on $x_R$, while $Y^\text{cl}_{LR}$ and $Y^\text{cl}_{RL}$ depend on pairs of eigenvalues with one on each side of the boundary cut (and $Y^{\text{cl}\, \dagger}_{RL} = Y^\text{cl}_{LR}$). Relatedly, the $\widetilde \Psi$ can also be partitioned by writing
\be\label{eq:psi}
\Psi = U \left(
\begin{array}{c}
    \widetilde \Psi_{L} \\
    \widetilde \Psi_{R}
\end{array}
\right) \,.
\ee

So far we have parametrized the physical phase space of the classical theory by $\{x, y,U,\widetilde \Psi\}$. These are subject to the additional constraints that $U U^\dagger = U^\dagger U = \text{Id}$ and $|\widetilde \Psi|^2 = N k$. We have shown how the target space partition of the eigenvalues into $x_L$ and $x_R$ can be uplifted into a block decomposition of the matrices after conjugation by $U$. This conjugation is, of course, a $U(N)$ gauge transformation. We must now look in more detail at the $U$ degrees of freedom themselves. The first point is that, when the system is considered in its entirety, the $U$ degrees of freedom do not have any dynamics associated to them. To see this we can write down the first order kinetic term in the Lagrangian, responsible for the quantum commutators between $X, Y$ and $\Psi$, in these variables:
\begin{align}
L_\text{kin} & = \frac{1}{2} \tr \left(Y \frac{d}{dt} X - X \frac{d}{dt} Y \right) + i \Psi^\dagger \frac{d}{dt} \Psi \label{eq:LZ} \\
& = i k \tr \left( U^\dagger \frac{d}{dt} U \right) + i \widetilde \Psi^\dagger \frac{d}{dt} \widetilde \Psi + y \frac{d}{dt} x \\
& = i k \frac{d}{dt} \left(\log \det U\right) + i \widetilde \Psi^\dagger \frac{d}{dt} \widetilde \Psi + y \frac{d}{dt} x \label{eq:tot} \\
& = i \hat \Psi^\dagger \frac{d}{dt} \hat \Psi + y \frac{d}{dt} x
\,. \label{eq:tot2}
\end{align}
Here we see that the $y$'s are just the conjugate momenta of the $x$'s. More importantly, however, in (\ref{eq:tot}) we see that the $U$'s appear inside a total time derivative. Thus, as expected, these pure gauge modes have no dynamics. However, we will now see that this is no longer the case once the system is partitioned. The partitioning will unleash `boundary cut' modes (our discussion here will be in the spirit of \cite{Donnelly:2016auv}) that are responsible for the gauge-theoretic entanglement of the partition. In the final line above we have shown how these non-dynamical gauge modes can naturally be absorbed as phases into the $\hat \Psi$, which we defined in (\ref{eq:psihat}). This step uses the fact that, with $X$ diagonalized, the diagonal components of the Gauss law (\ref{eq:gauss}) lead to the component-wise constraint that $|\widetilde \Psi_a|^2 = k$.

The unitary matrices themselves can be written in generality as
\be\label{eq:U}
U = V \left(
\begin{array}{cc}
    U_L & 0 \\
    0 & U_R
\end{array}
\right) \,.
\ee
If there are $M$ eigenvalues in $x_R$, then $U_L \in U(N-M)$ acts on an $N-M$ dimensional subspace, $U_R \in U(M)$ acts on the orthogonal $M$ dimensional subspace and $V$ parametrizes the quotient manifold $U(N)/[U(M) \times U(N-M)]$. The first part of our proposal is that the boundary cut modes do not include the $V$ degrees of freedom. That is, we will `gauge-fix'
\be
V = 1 \,.
\ee
It is clear that the subset of the unitary matrices with $V=1$ respect the block partition of matrices that we have made above, while a nonzero $V$ would mix the different blocks. Physically, and as we discuss further in section \ref{sec:moyal} below, the $V$ transformations include diffeomorphisms that mix the two regions of the bulk geometric partition that we are after (and a similar freezing of such modes was previously discussed in e.g.~\cite{Donnelly:2016auv}). The $U_L$ and $U_R$, in contrast, act on either side of the partition and are analogous to the gauge transformations that one has in a conventional lattice gauge theory setting. We will make some brief comments about re-instating the $V$'s later, but proceed to ignore them for the moment. That is to say, we will define a partition that leads to nontrivial phase space dynamics for $U_L$ and $U_R$, but not for $V$. This is consistent, as none of the $U$ degrees of freedom have any intrinsic dynamics to begin with.

With the above structure at hand, it is useful to introduce a `covariant' projection matrix that can be used to project onto the various blocks appearing in the partitions (\ref{eq:X}), (\ref{eq:Y}) and (\ref{eq:psi}). We define
\be\label{eq:Theta}
\Theta \equiv U  
\left(
\begin{array}{cc}
    1 & 0 \\
    0 & 0
\end{array}
\right) U^\dagger = \left(
\begin{array}{cc}
    1 & 0 \\
    0 & 0
\end{array}
\right) \,.
\ee
The last step holds because we have set $V=1$ in (\ref{eq:U}). Note that this projection keeps the lowest
$M$ eigenvalues, rather than those eigenvalues that are less than some critical value $x_o$. These are not the same in general because, for example, a given eigenvalue could fluctuate across the value $x_o$ which would require a jump in the value of $M$. However, we saw in the previous section that at large $N$ the distribution of eigenvalues is strongly peaked on the Wigner semicircle. At large $N$, then, there is a one-to-one correspondence between $M$ and $x_o$. Beyond large $N$ there will be a variance $\langle (\delta M)^2 \rangle$ corresponding to a fixed $x_o$. This small variance\footnote{Specifically: if $\delta n(x)$ is the fluctuation about the Wigner semicircle $n_o(x)$, then the number of surplus eigenvalues on one side of the cut is $\delta M = -\int_{-R}^{x_o} \delta n(x) dx$. 
Using the wavefunction (\ref{eq:cos}) for $\delta n(x)$, with the cosine modes cut off at $m=N$ per our discussion below equation (\ref{eq:sn}), one finds $\langle (\delta M)^2 \rangle \sim (\log N)/k \ll M^2$. We will always take $M$ to scale like $N$. Quantum fluctuations in the number of eigenvalues on each side of the cut are indeed small. \label{eq:Mfluc}
} will give a subleading logarithmic correction to the entanglement, as we discuss further in section \ref{sec:topological} below.

Using the projector (\ref{eq:Theta}) we can re-write the kinetic term in the Lagrangian (\ref{eq:LZ}) as
\be\label{eq:Lpart}
L_\text{kin} = i \tr \left(\Theta Z^\dagger \frac{d}{dt} Z + (1 - \Theta) Z^\dagger \frac{d}{dt} Z \right) + i \Psi_L^\dagger \frac{d}{dt} \Psi_L + i \Psi_R^\dagger \frac{d}{dt} \Psi_R \,.
\ee
Breaking up the trace in this way is analogous to writing e.g.~the Chern-Simons action as an integral over part of the space plus the integral over the complement. Recall that $Z$ was defined below equation (\ref{eq:H}) above. Note also that $Z^\dagger \frac{d}{dt} Z$ differs from $Y \frac{d}{dt} X - X \frac{d}{dt} Y$ by a total time derivative. This total derivative has been split between the two first terms in (\ref{eq:Lpart}), which will be important. We further defined $\Psi_L \equiv \Theta \Psi$ and $\Psi_R \equiv (1- \Theta) \Psi$. The Lagrangian (\ref{eq:Lpart}) can now be split as $L_\text{kin} = L_{L\,\text{kin}} + L_{R\,\text{kin}}$, with
\be
L_{L\,\text{kin}} = i \tr \left(\Theta Z^\dagger \frac{d}{dt} Z \right) + i \Psi_L^\dagger \frac{d}{dt} \Psi_L \,,
\ee
and similarly for $L_{R\,\text{kin}}$. The idea is that the left observer has access to the phase space described by $L_{L\,\text{kin}}$ while the right observer has access to the phase space described by $L_{R\,\text{kin}}$. We will see, however, that there are gauge-theoretic degrees of freedom (i.e. $U$'s) that are now nontrivial and which furthermore appear in both phase spaces (cf. \cite{Donnelly:2016auv}). These are the boundary cut modes. In the analogous Chern-Simons case, these would be gauge transformations that act nontrivially on the shared boundary of the two regions. The boundary cut modes must be identified between the two sides, and that will be the source of the gauge-theoretic entanglement.

Writing $Z$ and $\Psi$ in terms of $\{x,y,U_L,U_R,\widetilde \Psi\}$ the left and right kinetic terms become
\begin{align}
L_{L\,\text{kin}}  & = y_L \frac{d}{dt} x_L + \frac{i}{2} \tr \left(  Y^\dagger_{RL} \frac{d}{dt} Y_{RL}  \right)  + i \widetilde \Psi_L^\dagger \frac{d}{dt} \widetilde \Psi_L + i k \frac{d}{dt} \left(\log \det U_L\right) \,, \label{eq:LL} \\
L_{R\,\text{kin}}  & = y_R \frac{d}{dt} x_R + \frac{i}{2} \tr \left(  Y^\dagger_{LR} \frac{d}{dt} Y_{LR}  \right)  + i \widetilde \Psi_R^\dagger \frac{d}{dt} \widetilde \Psi_R + i k \frac{d}{dt} \left(\log \det U_R\right) \,, \label{eq:LR}
\end{align}
To obtain these expressions we have used the Gauss law projected onto each side of the split, i.e.~$\Theta G \Theta = 0$ and $(1-\Theta)G(1-\Theta) = 0$, dropped some terms that are total time derivatives, and set
\be
Y_{LR}^\dagger= Y_{RL} = U_R Y^\text{cl}_{RL} U_L^\dagger \,, \qquad \Psi_L = U_L \widetilde \Psi_L \,, \qquad \Psi_R = U_R \widetilde \Psi_R \,. \label{eq:relate}
\ee
In each of (\ref{eq:LL}) and (\ref{eq:LR}) the final two terms can be combined into a single term involving $\hat \Psi$, as previously in (\ref{eq:tot2}). As both the phase space and the ground state wavefunction factorize in terms of $\hat \Psi_L$ and $\hat \Psi_R$, these degrees of freedom do not contribute to the entanglement.
We saw in (\ref{eq:gshat}) that this split furthermore encompasses the overall $(\det U)^{k-1}$ term in (\ref{eq:gs}), which therefore also does not contribute to the entanglement. Note that upon summing (\ref{eq:LL}) and (\ref{eq:LR}) the second terms in each line add to give a total derivative. Upon ignoring this total derivative term, the full kinetic term recovers the previous result (\ref{eq:tot}). 

However, in either the left or right region, the $Y_{LR}$ modes are physical. These modes appear in both regions and we will identify them as the boundary cut modes. Connecting back to Fig.~\ref{fig:circleplot} these are off-diagonal modes associated to pairs of eigenvalues of $X$ on opposite sides of the entanglement cut. They transform as $Y_{LR} \to \Omega_L Y_{LR} \Omega^\dagger_{R}$ with $\Omega_L \times \Omega_{R} \in U(N-M) \times U(M)$ and are analogous to Wilson lines that cross the entanglement cut, as was emphasized in \cite{Hampapura:2020hfg}. We will flesh out that analogy in the following section.

To understand the quantum phase space of the $Y_{LR}$ we need their quantum commutation relations. These commutation relations do not follow immediately from e.g. (\ref{eq:LL}) because these variables inherit constraints from the unitarity of $U_L$ and $U_R$. These constraints can be dealt with as we now describe.

\subsection{The boundary cut modes}

We now proceed to characterize the Hilbert space in which the $Y_{RL}$ modes live upon quantization. We start by performing a change of variables at the classical level,  writing the singular value decomposition
\be\label{eq:singular}
Y_{RL} = \sum_{m=1} \lambda^{(m)} \xi_R^{(m)} \xi_L^{(m)\dagger} \,.
\ee
Here the $\xi_R^{(m)} \in \C^{M}$ and $\xi_L^{(m)} \in \C^{N-M}$ are two separate sets of orthonormal vectors, i.e. 
\be\label{eq:conss}
\xi_R^{(m)\dagger} \cdot \xi_R^{(n)} = \delta^{mn} \,, \qquad \xi_L^{(m)\dagger} \cdot \xi_L^{(n)} = \delta^{mn} \,,
\ee
and the $\lambda^{(m)}$ are the singular values. The sum over $m$ runs up to $\min(M,N-M)$. This decomposition introduces a gauge redundancy in which, for each $m$ separately, $\xi_R^{(m)}$ and $\xi_L^{(m)}$ are rotated by the same overall phase. The state that we will write down shortly for these modes will be invariant under this gauge symmetry.

The kinetic Lagrangian for the left boundary cut modes then becomes
\be\label{eq:LLbdy}
L_{L\, \text{bdy}} = \frac{i}{2} \tr \left(  Y^\dagger_{RL} \frac{d}{dt} Y_{RL}  \right) = \frac{i}{2} \sum_m \big|\lambda^{(m)}\big|^2 \left(\xi^{(m) \dagger}_R \cdot \frac{d}{dt} \xi_R^{(m)} + \xi_L^{(m)} \cdot \frac{d}{dt} \xi_L^{(m) \dagger} \right) \,.
\ee
Note that both $\xi_R$ and $\xi_L$ appear in the left (and in the right) hand Lagrangian. The singular value decomposition (\ref{eq:singular}) is analogous, as we see in more detail and the end of this subsection, to writing the gauge connection across the cut as a difference of phases (contained within $\xi_R$ and $\xi_L^\dagger$ in this case). The boundary mode, that is to be duplicated, is the connection itself, which is written in terms of both phases. Similarly, upon duplicating $Y_{RL}$, both $\xi_R$ and $\xi_L$ will be duplicated. One should not be confused by the notation which may suggest (incorrectly) that $\xi_R$ is only assigned to the right and $\xi_L$ to the left.

We can decompose $Y_{RL}^\text{cl}$ in the same way and obtain, using (\ref{eq:relate}), the relations
\be\label{eq:rotate}
\lambda^{(m)} = \lambda^{(m)}_\text{cl} \,, \qquad \xi_R^{(m)} = U_R \xi_{\text{cl}R}^{(m)}
\,, \qquad \xi_L^{(m)} = U_L \xi_{\text{cl}L}^{(m)} \,.
\ee
Recall from (\ref{eq:yclass}) that $Y_{RL}^\text{cl}$, and hence $\{\lambda^{(m)}_\text{cl}, \xi_{\text{cl}R}^{(m)}, \xi_{\text{cl}L}^{(m)}\}$, are given in terms of the eigenvalues $x$ and vector $\widetilde \Psi$. They are independent of $U$. As we explain around (\ref{eq:tt}) below, the components of $\widetilde \Psi$ are all fixed by Gauss's law, up to phases that drop out of $Y_{RL}^\text{cl}$. Furthermore, we have seen that at large $N$ the eigenvalue distribution is strongly peaked at its classical value. We will discuss the subleading effects of quantum fluctuations of the eigenvalues, and hence of $\lambda_\text{cl}$, later in section \ref{sec:topological}. Meanwhile, at large $N$ we will fix $\lambda_\text{cl}$ to its classical value in the kinetic term (\ref{eq:LLbdy}), which then describes the dynamics of $U_L$ and $U_R$. These unitary matrices are conveniently packaged by (\ref{eq:rotate}) into the $\xi^{(m)}_R$ and $\xi^{(m)}_L$. We can think of (for example) the $\xi_{\text{cl}R}^{(m)}$ as a given fixed frame of orthonormal vectors. According to (\ref{eq:rotate}), the matrix $U_R$ describes all possible rotations of this frame into a new frame $\xi^{(m)}_R$. We wish to quantize the phase space of these rotated frames.

From (\ref{eq:LLbdy}) the rotated `frame vectors' are raising and lowering operators obeying the commutation relations
\be\label{eq:comm}
\left[\xi_{Ra}^{(m)}, \xi_{Rb}^{(n)\dagger}\right] = \frac{2}{\big|\lambda^{(m)}_\text{cl}\big|^2} \delta_{ab} \delta^{mn} \,, \qquad \left[\xi_{La}^{(m)\dagger}, \xi_{Lb}^{(n)}\right] = \frac{2}{\big|\lambda^{(m)}_\text{cl}\big|^2} \delta_{ab} \delta^{mn} \,.
\ee
Furthermore, using the commutators (\ref{eq:comm}) the constraints (\ref{eq:conss}) commute with each other on the constrained subspace and with the Hamiltonian\footnote{The Hamiltonian for each region, which descends purely from the potential term (\ref{eq:H}), doesn't depend on $U_L$ or $U_R$.} --- they are first class constraints and do not modify the commutation relations. The constraints (\ref{eq:conss}) are to be imposed directly on the Fock space of quantum states generated by the $\xi_R^{(m)}$ and $\xi_L^{(m)}$ with commutators (\ref{eq:comm}).

There is one more important point that concerns the distinguishability of the quanta created by the various $\xi_R^{(m)\dagger}$. In sections \ref{sec:model} and \ref{sec:col} we have emphasized that the classical collective field is built out of ordered eigenvalues. Equivalently, the collective field depends on the number of eigenvalues with a given value but not on which eigenvalues these are. Ordering eliminates this redundancy and was achieved above via an $S_N$ gauge transformation. Furthermore, when we parameterized $X$ by its eigenvalues in (\ref{eq:X}), the eigenvalues were taken to be ordered. This was important for the projection matrix $\Theta$ defined by (\ref{eq:Theta}) to correspond to a geometric partition.
However, as written, the relations (\ref{eq:rotate}) re-introduce permutations of the eigenvalues on each side of the cut as physical operators, as the $U_{L/R}$ matrices include permutations. Because we are looking for boundary cut modes that are built upon the emergent large $N$ geometry of the collective field, we should not include these permutations. Thus we restrict
\be\label{eq:nop}
U_L \in U(N-M)/S_{N-M} \,, \qquad U_R \in U(M)/S_{M} \,.
\ee
We are free to do this for a similar reason that we were free to set $V=1$ previously: a priori none of the gauge degrees of freedom are physical, but they can become physical due to the nature of the entanglement cut. However, the choice of cut includes a choice of which gauge modes are `resurrected' in this manner. Our choice here is guided by the nature of the classical collective field.

Before moving on let us make a connection between the $Y_{RL}$ and the more conventional boundary cut modes of a $U(1)$ lattice gauge theory. A related discussion has been given in \cite{Hampapura:2020hfg}. As we have already noted, under a gauge transformation $Y_{RL} \to \Omega_R Y_{RL} \Omega_L^\dagger$. Consider the particular unitary matrices given by
\be\label{eq:Omega}
\Omega^{(m)}_R = e^{i \theta^{(m)}_R} \xi^{(m)}_R \xi^{(m)\dagger}_R \,, \qquad \Omega^{(m)}_L = e^{i \theta^{(m)}_L} \xi^{(m)}_L \xi^{(m)\dagger}_L \,.
\ee
Here $\theta^{(m)}_R$ and $\theta^{(m)}_L$ are phases. These unitaries generate the gauge transformation
\be
Y_{RL} \to e^{i \left(\theta^{(m)}_R - \theta^{(m)}_L\right)} Y_{RL} \,.
\ee
This is precisely the transformation of a $U(1)$ Wilson line connecting two vertices of a lattice. In \ref{sec:moyal} below
we will elaborate on the geometric interpretation of this transformation: the projectors $\xi^{(m)}_R \xi^{(m)\dagger}_R$
and $\xi^{(m)}_L \xi^{(m)\dagger}_L$ are associated to geometric regions on opposite sides of the cut, while the $Y_{RL}$ are Wilson lines connecting these regions. The strength of the connectivity between the two regions in the quantum state will be determined by the corresponding singular value $\lambda^{(m)}_\text{cl}$.

\subsection{Singular values of the `off-diagonal' elements}
\label{sec:singular}

The upshot of the previous section is that, on each side of the cut, the boundary cut Hilbert space is the Fock space built using the raising and lowering operators (\ref{eq:comm}) and subject to the constraints (\ref{eq:conss}). Furthermore, (\ref{eq:nop}) requires that we restrict to permutation symmetric states because the permutation subgroups are, by fiat, not part of the boundary cut degrees of freedom.

We must now write down a gauge-invariant quantum state for these boundary modes. 
To understand this state we must first get a handle on the singular values $\lambda^{(m)}_\text{cl}$. We will see in the remaining sections that these singular values are at the crux of the connection between geometry and entanglement in our model.
In this section we will show that the matrix $Y_{RL}^\text{cl}$ is low rank: almost all of the singular values are exponentially small. This means that, effectively, many of the $U_R$ and $U_L$ matrices do not in fact act on $Y_{RL}^\text{cl}$ and will not contribute to the entanglement of a given cut. That is, the emergent lattice gauge theory description has a low lattice connectivity. This will simplify writing down the boundary mode quantum state and evaluating its entanglement entropy.

The components of $Y_{RL}^\text{cl}$ are given by the lower diagonal entries of (\ref{eq:yclass}). That is,
\be\label{eq:classY}
\left(Y^\text{cl}_{RL}\right)_{ab} = - i \frac{\Theta(x_a - x_o)\Theta(x_o - x_b)}{x_a - x_b}  \widetilde \Psi_b \widetilde \Psi^\dagger_a \,.
\ee
This matrix can be simplified by recalling, from below (\ref{eq:tot}), that the diagonal components of the Gauss law (\ref{eq:gauss}), with $X$ diagonalized, lead to the component-wise constraint that $|\widetilde \Psi_a|^2 = k$. Therefore $\widetilde \Psi_a = \sqrt{k} e^{i \vartheta_a}$, for some phase $\vartheta_a$. These phases will cancel out when we compute traces of $Y^\text{cl}_{RL}$. It follows that in traces we may set $\widetilde \Psi_b \widetilde \Psi^\dagger_a \to k$ in (\ref{eq:classY}). In particular, the singular values that we are interested in can be extracted from the following traces, with $s$ a positive integer,
\begin{align}
\sum_m \left|\lambda^{(m)}_\text{cl} \right|^{2s} & = \tr\left[ \left( Y^{\text{cl}\,\dagger}_{RL} Y^\text{cl}_{RL} \right)^s \right] \label{eq:micro} \\
 & = k^{2s} \int^{x_o} d^sx \int_{x_o} d^sy  \frac{n(x_1) n(y_1) \cdots n(x_s) n(y_s)}{(y_1 - x_1)(y_1-x_2)(y_2-x_2)(y_2-x_3) \cdots (y_s-x_s)(y_s-x_1)} \,.\label{eq:tt}
 \end{align}
In the second line we have expressed the sum over eigenvalues $x_a$ in terms of an integral over the eigenvalue distribution $n(x)$ from section \ref{sec:col}.
We proceed to evaluate these integrals and extract the values of $\left|\lambda^{(m)}_\text{cl} \right|$ from the result.

The integral (\ref{eq:tt}) is logarithmically divergent due to contributions from $y_i \sim x_i \sim x_o$. This divergence is an artifact of using the collective field to perform the traces and is regulated by the graininess of the eigenvalue description. It is sufficient for our purposes to isolate the logarithmically divergent part, as this will dominate the integral at large $N$. To do this one may set $y_i = x_o + \Delta y_i$ and $x_i = x_o + \Delta x_i$ and expand the integrand for small $\Delta y_i$ and $\Delta x_i$. The net effect is that all of the $n(x)$ and $n(y)$ in (\ref{eq:tt}) become $n(x_o)$. Furthermore performing the $\Delta y$ integrals one obtains
\begin{align}
\sum_m \left|\lambda^{(m)}_\text{cl} \right|^{2s} & = k^{2s} n(x_o)^{2s} \int_{-\infty}^0 d^s \Delta x \frac{\log \frac{\Delta x_2}{\Delta x_1} \log \frac{\Delta x_3}{\Delta x_2} \cdots \log \frac{\Delta x_1}{\Delta x_s}}{(\Delta x_1 - \Delta x_2) (\Delta x_2 - \Delta x_3) \cdots (\Delta x_s - \Delta x_1)} \label{eq:fi} \\
& = k^{2s} n(x_o)^{2s} \frac{(2 \pi)^{2s-2} \Gamma(s)^2}{\Gamma(2s)} \int_{-\infty}^0 \frac{d \Delta x_1}{\Delta x_1} \label{eq:div}\\
& = k^{2s} n(x_o)^{2s} \frac{(2 \pi)^{2s-2} \Gamma(s)^2}{\Gamma(2s)} \log \frac{\hat R}{\hat \epsilon} \label{eq:thr} \\
& = \int_0^{\lambda_o} d\lambda \lambda^{2s} \frac{\log \frac{\hat R}{\hat \epsilon}}{\pi^2 \lambda \sqrt{1-\l^2/\lambda_o^2}} \,. \label{eq:fin}
\end{align}
In the third line, the logarithmic divergence is cut off at small distance by the inter-eigenvalue spacing $\hat \epsilon$. Corrections to the above expressions from higher order terms in the expansion about $x_o$ give additional factors of $\Delta x$ in the numerator and are therefore not divergent. The value $\hat R$ in the logarithm cannot be determined by this approach, but is expected to be
set by the extent of the eigenvalue distribution. The cutoff $\hat \epsilon$ is not, a priori, the same as the $\epsilon$ cutoff discussed around equation (\ref{eq:ent}) above. It is clear from (\ref{eq:div}) that $\hat \epsilon$ is the distance between eigenvalues, so that $\hat R/\hat \epsilon \sim R n(x_o) \sim N \sqrt{1 - x_o^2/R^2}$.

In the final line (\ref{eq:fin}) we wrote the answer in the form of a Mellin transform with
\be\label{eq:lam0}
\lambda_o = k n(x_o) \pi \,.
\ee
From this expression we can read off the distribution $\rho(\lambda)$ of (the modulus of the) singular values of $Y_{RL}^\text{cl}$:
\be\label{eq:rholam}
\rho(\lambda) = \frac{\Theta(\lambda_o - \lambda) \log \frac{\hat R}{\hat \epsilon}}{\pi^2 \lambda \sqrt{1-\l^2/\lambda_o^2}} \,.
\ee
Here $\Theta$ is the Heaviside step function.
This distribution should not be confused with the distribution $n(x)$ of the eigenvalues of $X$. The appearance of a semi-circle-like factor $\sqrt{1-\l^2/\lambda_o^2}$ in (\ref{eq:rholam}) is suggestive of an underlying randomness in the $Y_{RL}$ matrix. Note, however, that this is distinct from the semicircle distribution of the eigenvalues of $X$.

We now show that the distribution function (\ref{eq:rholam}) corresponds to singular values
$|\lambda^{(m)}_\text{cl}|$ that decay exponentially with $m$. The distribution is related to the singular values themselves through $\rho(\lambda) = dm/d|\lambda^{(m)}_\text{cl}|$. Integrating (\ref{eq:rholam}) and inverting one finds
\be\label{eq:lamcl}
|\lambda^{(m)}_\text{cl}| = \lambda_o \sech \frac{\pi^2 m}{\log \frac{\hat R}{\hat \epsilon}} \,.
\ee
Thus the first singular value $|\lambda^{(0)}_\text{cl}| = \lambda_o$ in (\ref{eq:lam0}), and the remainder decay exponentially so that there are of order $\log \frac{\hat R}{\hat \epsilon} \sim \log N$ effectively nonzero singular values. The number of these nonzero values is small compared to the total number of singular values: $\log N \ll \min\{M,N-M\} \sim N$ (we will always take $M$ to be of order $N$). It follows that the matrix $Y_{LR}^\text{cl}$ is indeed low rank. 

In the above discussions we are fixing the eigenvalue distribution to the classical Wigner distribution $n_o(x)$ in (\ref{eq:wigner}). The correction due to fluctuations can be estimated using the Gaussian wavefunction (\ref{eq:cos}) to obtain $\langle \delta n(x_o) \delta n(x_o + 0^+)\rangle/n_o(x_o)^2 \sim 1/(k N)$. It is important to split the points slightly in evaluating the quantum fluctuation to avoid UV divergences (which are regulated by $N$). In the original microsopic expression (\ref{eq:micro}) all points are indeed distinct. In the estimate we further used $n_o(x_o) \sim \sqrt{N/k}$ from (\ref{eq:wigner}).
It follows that quantum eigenvalue fluctuations can
be neglected in calculating the entanglement entropy to leading order. Quantum fluctuations of the singular values can, however, lead to a subleading logarithmic contribution to the entanglement entropy that we discuss further in section \ref{sec:topological} below.

We have verified the above expressions numerically, as described in Appendix \ref{sec:num}. One subtle point is that while the $x_o$ dependence of e.g.~(\ref{eq:thr}) is robustly reproduced, that is to say $\tr\left[ \left( Y^{\text{cl}\,\dagger}_{RL} Y^\text{cl}_{RL} \right)^s \right] \propto n(x_o)^{2s} \log n(x_o)$, the coefficient of proportionality converges logarithmically slowly as $N \to \infty$.

\subsection{State of the boundary cut modes}
\label{sec:state}

Consider first the simplified situation that there is only one nonzero singular value. This will be the case for the circular cut in section \ref{sec:circular} below. For the vertical cut we have seen that there are very few non-negligible singular values. The vast majority of the $\xi$ oscillators correspond to exponentially small singular values and will make exponentially small contributions to the entanglement entropy. This suggests that the case of a single nonzero singular value should be a helpful starting point.

We firstly write down a state in the Fock space of the boundary cut modes that is invariant under $U(M)$ and $U(N-M)$ and that solves the diagonal ($m=n$) constraint in (\ref{eq:conss}). With only a single $\xi$ vector, there are no orthogonality ($m \neq n$) constraints, which simplifies things. As we have stressed below (\ref{eq:LLbdy}), both $\xi_R$ and $\xi_L$ are observables on both sides of the cut. This is because the boundary modes $Y_{RL} = Y_{LR}^\dagger$, that appear in both the left and right Lagrangians (\ref{eq:LL}) and (\ref{eq:LR}), contain both $U_L$ and $U_R$. If we denote $\xi_R$ and $\xi_L$ as the modes on a given side of the cut, there is then a second copy, which we denote $\eta_R$ and $\eta_L$, that are observables on the other side of the cut. With this structure at hand we can write down the state
\be\label{eq:cutstate}
| \psi_\text{cut} \rangle' = \left(\xi_R^\dagger \cdot \eta_R^\dagger \right)^{|\lambda_\text{cl}|^2/2} \Big(\xi_L \cdot \eta_L \Big)^{|\lambda_\text{cl}|^2/2} |0\rangle \,.
\ee
Note that, according to (\ref{eq:comm}), the raising operators are $\xi_R^\dagger$ and $\xi_L$. In order to write down this $U(M)\times U(N-M)$ invariant state it was essential to have the two copies $\xi$ and $\eta$ of the oscillators. However, these two copies are strongly entangled by the state (\ref{eq:cutstate}) and this will lead to the gauge-theoretic entanglement that we compute shortly. We can also note that (\ref{eq:cutstate})
is invariant under the gauge symmetry introduced by the singular value decomposition, noted below (\ref{eq:singular}), in which both $\xi_R$ and $\xi_L$ are rotated by the same $m$-dependent phase.

We can now observe that (\ref{eq:cutstate}) describes a maximally entangled state between the $\xi$ and $\eta$ oscillators. Taking for concreteness $\xi_R$ and $\eta_R$, that are $M$-component vectors, the multinomial expansion gives (asymptotically we can assume that $|\lambda_\text{cl}|^2/2 \equiv p$ is an integer)
\begin{equation}\label{eq:eqn1}
    (\xi_R^{\dag} \cdot \eta_R^{\dag})^{p}|0\rangle = \sum_{p_1 + p_2 + \cdots + p_M =p}\frac{p!}{p_1!p_2!\ldots p_M!}\prod_{a=1}^M (\xi_{Ra}^{\dag} \eta_{Ra}^{\dag})^{p_a}|0\rangle \,.
\end{equation}
Using the oscillator algebra $a^{\dag q}|0\rangle = \sqrt{q!}|q\rangle$ we can denote $(\xi_a^{\dag} \eta_a^{\dag})^{p_a}|0\rangle = p_a! |p_a\rangle_\xi|p_a\rangle_\eta$. These factorials cancel the denominator of (\ref{eq:eqn1}) so that we obtain
\begin{equation}\label{eq:eqn2}
    (\xi_R^{\dag} \cdot \eta_R^{\dag})^{p}|0\rangle = \sum_{p_1 + p_2 + \cdots + p_M = p} p! |p_1,p_2,\ldots,p_M\rangle_{\xi_R} |p_1,p_2,\ldots,p_M\rangle_{\eta_R} \,.
\end{equation}
This describes a maximally entangled state on the Fock subspace with total occupation number $p$ for each of the $\xi$ and $\eta$ oscillators. The state (\ref{eq:eqn2}) is then to be tensored with the analogous state for $\xi_L$ and $\eta_L$, in which $M \leftrightarrow N-M$.

Upon tracing out the $\eta$ oscillators, the maximally entangled state (\ref{eq:eqn2}) gives a density matrix that is proportional to the identity on a subspace of the Hilbert space. Such a density matrix is, consistently with our starting point (\ref{eq:cutstate}), automatically invariant under the action of $U(M)$ and $U(N-M)$. The gauge invariance of maximally entangled states allows us to take the next step, which is to account for the quotient by permutations in (\ref{eq:nop}). With only a single $\xi_R$ and $\xi_L$ vector, this quotient implies that the $M$ components of $\xi_R$ and $N-M$ components of $\xi_L$ are indistinguishable. It is therefore natural to write down the maximally entangled state obtained by projecting (\ref{eq:eqn2})
to the symmetric polynomials of creation operators. Without keeping track of the overall normalization of the state we can write this as
\be\label{eq:quota}
| \psi_R \rangle = \sum_{\sum p_a = p} |p_1,p_2,\ldots,p_M\rangle_{\xi_R S} |p_1,p_2,\ldots,p_M\rangle_{\eta_R S} \,.
\ee
The $S$ subscript denotes symmetrization over the labels of the state (as for indistinguishable particles). That is to say, the labels of $|p_1,p_2,\ldots,p_M\rangle_{S}$ can be taken to be ordered. The full boundary cut state is then
\be\label{eq:pcut}
|\psi_\text{cut}\rangle = | \psi_R \rangle| \psi_L \rangle \,.
\ee
Here the state $| \psi_L \rangle$ for the $\xi_L$ and $\eta_L$ modes is as in (\ref{eq:quota}), but with $M \leftrightarrow N - M$. It is important to appreciate that, due to the quotient by permutations, the state (\ref{eq:quota}) is maximally entangled over a much smaller subspace than (\ref{eq:eqn2}). This fact will be crucial in order to obtain an area law entanglement shortly.

Moving beyond a single singular value leads to two complications. Firstly, the various $\xi^{(m)}$ vectors must obey orthogonality constraints, which are the ($m \neq n$) constraints in (\ref{eq:conss}). Secondly, according to (\ref{eq:rotate}) the gauge rotations act on the entire frame and therefore, in particular, the permutations act simultaneously on all of the $\xi^{(m)}_R$ vectors. For example, if the permutation swaps the first two components of $\xi^{(1)}_R$ then it also swaps the first two components of all of the remaining $\xi^{(m)}_R$. We will now argue that, because $Y_{RL}$ is effectively low rank, to leading order at large $N$ and $M$ these two complications conspire to cancel each other out so that the state may be taken to be
\be\label{eq:cutfull}
|\psi_\text{cut}\rangle \approx \prod_m |\psi^{(m)}_\text{cut}\rangle \,.
\ee
Here $|\psi^{(m)}_\text{cut}\rangle$ is just the state written previously in (\ref{eq:pcut}) with $p \to |\lambda^{(m)}|^2/2$, $\xi \to \xi^{(m)}$ and $\eta \to \eta^{(m)}$. Let us now justify (\ref{eq:cutfull}).

Consider all pairs of classical $M$-component vectors $\xi^{(1)}$ and $\xi^{(2)}$ that are orthogonal. There is a preferred subset of these orthogonal pairs insofar as the action of $S_M$ is concerned. Recall that $S_M$ acts on the components of these vectors, in the basis in which $X$ was diagonalized. One can consider vectors such that $\xi^{(1)}$ has only $M^{(1)}$ nonzero components in this basis, while $\xi^{(2)}$ is only nonzero on the complementary $M^{(2)} = M - M^{(1)}$ components.
What is special about such pairs of orthogonal vectors is that the nontrivial action of $S_M$ splits into an action of $S_{M^{(1)}} \times S_{M^{(2)}}$. These smaller permutation groups act independently on the two vectors. The additional elements of $S_M/[S_{M^{(1)}} \times S_{M^{(2)}}]$ swap around which components of $\xi^{(1)}$ are nonzero. We can gauge-fix this part of the symmetry by taking the first $M^{(1)}$ components of $\xi^{(1)}$ to be nonzero and then the remaining $M^{(2)}$ components of $\xi^{(2)}$ to be nonzero. There is a physical choice here to reduce the space of boundary cut modes
\be
U(M)/S_M \to U(M^{(1)})/S_{M^{(1)}} \times U(M^{(2)})/S_{M^{(2)}} \,.
\ee
This is an additional restriction beyond that described already in (\ref{eq:nop}). In section \ref{sec:annulus} below we will see an especially transparent case of this reduction, where the two nontrivial singular values are associated to the inner and outer boundary of an annulus. More generally, the reduction is suggested by the orthogonality constraint. Clearly, the construction we have just described can be generalized to a number $P$ of vectors by writing $M = M^{(1)} + M^{(2)} + \cdots + M^{(P)}$. Quantizing this smaller phase space one can write down the state (\ref{eq:cutfull}) which now
has vectors of length $M^{(1)}$, $M^{(2)}$, $\ldots$, $M^{(P)}$ and obeys all constraints. We propose that this is the state of the boundary cut modes in the case of multiple nonzero singular values.

There is some imprecision in the above construction: We need to specify how many of the $\xi^{(m)}$ should be kept and also what values should be chosen for $M^{(m)}$. Again let $P$ denote the number of vectors $\xi^{(m)}$ that are considered to correspond to `nonzero' singular values $\lambda_\text{cl}^{(m)}$. We will now explain that as long as $P$ is sufficiently small, the different possible choices of partition $M = \sum_{m=1}^P M^{(m)}$ give a subleading correction to the entanglement entropy at large $M \sim N$. Furthermore, the leading order entanglement entropy will be given shortly as a sum over the singular values, and hence
truncating the singular values at different values of $P$ will lead to only exponentially small changes to the entanglement entropy.

Suppose for example that $P=2$. The number of ways we can partition $M$ into two integers is of order $M$. For a typical partition, $M^{(1)}$ and $M^{(2)}$ are both also of order $M$. The number of ways that we can solve the diagonal constraint $\sum_{a=1}^{M^{(1)}} p_a = \frac{1}{2} |\lambda_\text{cl}^{(1)}|^2$ by choosing ordered integers $p_a$ is, in contrast, exponentially large in $\lambda^{(1)}_\text{cl}$. This counting will be discussed in more detail in the following section where we will recall that 
$\lambda^{(1)}_\text{cl}$ grows with $M \sim N$. It follows that the uncertainty in the partition of $M$ is tiny compared to the uncertainty in the partition of $|\lambda_\text{cl}^{(1)}|^2$. The uncertainty in how $M$ is partitioned can lead at most to a subleading logarithmic correction to the entanglement entropy, of the kind that will be discussed in section \ref{sec:topological}.

We can now compute the entanglement between the $\xi$ and $\eta$ modes in the state (\ref{eq:cutfull}).

\subsection{The gauge-theoretic entanglement}
\label{sec:gauge}

The state (\ref{eq:cutfull}) is a product of maximally entangled states. The entanglement entropy is the sum of the entanglement entropy of each factor, which is just the logarithm of the dimension of the Hilbert space $\mathcal{H}^{(m)}_S$ in which that factor lives. That is
\be\label{eq:scut1}
s_\text{cut} = 2 \sum_m \log \text{dim} \mathcal{H}^{(m)}_S \,.
\ee
The factor of $2$ is from the left and right oscillators ($\xi_L$ and $\xi_R$). We will see shortly that the difference of $M$ vs $N-M$ between these is not important.

Because the labels in the maximally entangled state (\ref{eq:quota}) are indistinguishable (equivalently, ordered), the dimensions in (\ref{eq:scut1}) are given by the number of ways $p = \frac{1}{2} |\lambda_\text{cl}^{(m)}|^2$ can be partitioned into the $M$ (or $N-M$) integers $p_a$. If the number of integers in the partition is large compared to $p$, specifically if
$\sqrt{p} \ll \{M, N - M\} \sim N$,
then the number of partitions is given asymptotically by the Hardy-Ramanujan formula without needing to worry about the maximal number ($M$ or $N-M$) of integers in the partition:
\be\label{eq:logdim}
\log \text{dim} \mathcal{H}^{(m)}_S = \frac{\pi |\lambda_\text{cl}^{(m)}|}{\sqrt{3}} - 2 \log |\lambda_\text{cl}^{(m)}| + \cdots \,.
\ee
The condition for use of the Hardy-Ramanujan formula is then seen to be, using (\ref{eq:lam0}) and (\ref{eq:wigner}) to get the scaling of $p$ with $N$ and $k$, that $k \ll N$. Thus, while we may take $k$ large if we wish, in that case the large $N$ limit must be taken first.\footnote{If a large $k$ limit is taken before the large $N$ limit one ends up in a non-geometric semiclassical regime where the entanglement is given by the volume of the classical phase space. The classical phase space associated to each singular value is a $2M$ dimensional sphere of radius $|\lambda_\text{cl}|$. The quotient by $S_M$ divides the volume of the sphere by $M!$. At large $M$, then, the volume ($\text{vol}$) is given by $\log \text{vol} \sim M \log \left(|\lambda_\text{cl}|^2/M^2 \right) \sim M \log (k/N)$ (taking $M \sim N$). Thus is $k \gg N$ the classical phase space volume is large and will give the entanglement, but if $k \ll N$ then the classical phase space volume is small and the more refined Hardy-Ramanujan counting is necessary.\label{largek}} In (\ref{eq:logdim}) we see that the entanglement is directly related to the singular values.

To perform the sum in (\ref{eq:scut1}) we must integrate the Hilbert space dimensions (\ref{eq:logdim}) against the distribution (\ref{eq:rholam}) of singular values. The result is dominated by the larger values of $\lambda_\text{cl}$ so that at large $\lambda_o$ it is okay to use the expansion (\ref{eq:logdim}) inside the integral, giving
\begin{align}
s_\text{cut} = \left[\frac{\l_o}{\sqrt{3}} - \frac{2}{\pi^2} (\log \l_o)^2 + \cdots \right] \log \frac{\hat R}{\hat \epsilon} \,. \label{eq:scut11}
\end{align}
To obtain the subleading logarithmic term we cut off the small $\lambda$ divergence in the integral at $\frac{1}{2}\lambda^2 \sim 1$ (corresponding to $p\sim 1$ in the partition formula). The precise numerical value of the cutoff does not affect the answer shown above at large $\lambda_o$.

We now focus on the leading term in (\ref{eq:scut11}). Above we have noted that there are various other potential contributions to the subleading logarithmic correction that we are not keeping track of here. We discuss these again in section \ref{sec:topological} below. For the leading term we can recall that $\lambda_o = k n(x_o) \pi $ according to (\ref{eq:lam0}). The key fact here is that $n(x_o)$, and hence the dominant singular value $\lambda_o$, is proportional to the length of the entanglement cut. More precisely,
using the semicircle distribution (\ref{eq:wigner}) gives
\begin{align}
s_\text{cut} = \frac{(Nk)^{1/2} \log(N L)}{\sqrt{6}} L + \cdots \,. \label{eq:sbulk}
\end{align}
Here the length $L$ of the entanglement cut was defined in (\ref{eq:len}). From the discussion below (\ref{eq:fin}) the cutoff $\log \frac{\hat R}{\hat \epsilon} = \log (NL)$, up to non-universal terms in the logarithm that do not have any singular dependence on $N$ or $L$.

The gauge-theoretic cut entanglement (\ref{eq:sbulk}) is seen to be proportional to the length of the cut through the bulk. That is, it is consistent with an emergent bulk locality. 
The logarithmic multiplicative factor of the `area law' in (\ref{eq:sbulk}) indicates a mild violation of locality in the bulk cutoff. We will see in the following section \ref{sec:circular} and in appendix \ref{app:square} that this violation is tied up with the fact that the cut intersects the boundary of the droplet. When the cut remains in the interior of the droplet there is precisely one singular value per connected component of the cut. We have seen above that the logarithm is technically due to the fact that there is a multiplicity --- indeed of order $\log (NL)$ --- singular values. Each individual singular value contributes an area law without a logarithmic factor.

The cut term (\ref{eq:sbulk}) is to be added to the boundary contribution (\ref{eq:ent}) that we found previously. As expected, the cut contribution is more strongly divergent as $N \to \infty$. The contributions may be discussed in a unified fashion, as previously explained in \cite{Hampapura:2020hfg}. The state of the boundary cut modes $\xi$ depends on the collective field $n(x)$. The total entanglement entropy, dependent on both the state of $n(x)$ and the state of the $\xi$, is therefore
\be\label{eq:ptot1}
s = - \sum_{n(x),\xi} p[n,\xi] \log p[n, \xi] = - \sum_{n(x)} p[n] \log p[n] + \sum_{n(x)} p[n] \log \text{dim} \mathcal{H}_n \,.
\ee
Here $p[n,\xi] = p[n]/\text{dim} \mathcal{H}_n$, where $p[n]$ is the probability distribution of the collective field $n(x)$ and $1/\text{dim} \mathcal{H}_n$ is the probability of each of the maximally entangled $\xi$ states, given the value of $n$.
The first term in (\ref{eq:ptot1}) is the entanglement in the collective field, this is precisely the boundary mode entanglement that we calculated in section \ref{sec:col}. The second term in (\ref{eq:ptot1}) is the Hardy-Ramanujan result (\ref{eq:logdim}) averaged over $n$. As we have discussed in section \ref{sec:singular} above, $p[n]$ is strongly peaked on the classical solution $n_o(x)$, with Gaussian fluctuations $\delta n$ determined by (\ref{eq:psi3}). Because the fluctuations are small at large $N$, this term is well-approximated by (\ref{eq:sbulk}).

It should be emphasized again that the permutation symmetry imposed in writing down the entangled state (\ref{eq:quota}) played an essential role in obtaining an `area law' entanglement. If all of the entangled oscillator states are kept, the entanglement is much greater. The symmetrization, and associated counting of states via the Hardy-Ramanujan formula, also gives a physical connection to the chiral boundary cut mode of Chern-Simons theory. In Chern-Simons theory the cut entanglement is due to a maximally entangled pair of chiral bosons propagating along the cut (e.g.~\cite{PhysRevLett.96.110404,Das:2015oha,Wong:2017pdm,Balasubramanian:2018por} for some representative perspectives). More precisely, the entanglement is given by the (regulated) high temperature thermodynamic entropy of a chiral boson. That is $s_\text{CS} = \lim_{T \to \infty} \pa_T (T \log Z)$ where the chiral boson partition function
\be\label{eq:sCS}
Z(T) = \sum_{p>0} P(p) e^{- (p + \frac{1}{2})/T} \,.
\ee
Here $P(p)$ are the same partitions of integers that appeared in our counting problem above, with asymptotics given by the Hardy-Ramanujan formula. In (\ref{eq:sCS}) we have omitted the subleading contribution from winding modes of the chiral boson, we return to those modes briefly in section \ref{sec:topological}. At high temperatures $T$ the sum in (\ref{eq:sCS}) can be evaluated by saddle point. Using the large $p$ growth $P(p) \sim e^{\pi \sqrt{2 p/3}}$ we see that the saddle point value is $p_\star = \pi^2 T^2/6$. In terms of this quantity, $s_\text{CS} = \pi \sqrt{2 p_\star/3}$. If we then set $p_\star = \frac{1}{2} |\lambda_\text{cl}|^2$ we obtain precisely the leading order term  that we found above in (\ref{eq:logdim}). Therefore, for each singular value, the counting problem we have set up for the off-diagonal matrix modes is equivalent to that of the chiral boson boundary cut modes of Chern-Simons theory.

\subsection{The circular cut}
\label{sec:circular}

We have focused on a vertical cut through the droplet, as the ground state wavefunction is especially simple in the corresponding variables. This allows control of the fully quantum state. However, it is also possible to cut the droplet in a circle at a fixed radius $r_o < R$. The entanglement is expected to be purely gauge theoretic in that case as the cut does not intersect the boundary of the droplet. We will now compute this entanglement using a different method to the vertical cut but obtaining a very similar result. This fact helps to build confidence in our computations. Furthermore, the circular cut (and a `square' cut that we consider in Appendix \ref{app:square}) will clarify the origin of the logarithmic violation of the area law that we found above.

A partition at fixed radius is defined using a projection matrix $\Theta_{I}$, with $I$ for `inside'. This projector is now taken to be a function of the matrix $R^2 = 2 Z^\dagger Z$ (which differs from $X^2+Y^2$ due to matrix noncommutativity), rather than $X$ as previously. Specifically, $\Theta_{I}$ projects to the lowest $M$ eigenvalues of $R$. Similar circular cuts have been discussed previously in \cite{Das:2020xoa}, but our semiclassical computation will be technically somewhat different. The ground state wavefunction is not known explicitly in terms of the eigenvalues of $R$. However, the classical ground state is especially simple in these variables. In the basis in which $R$ is classically diagonal, the classical ground state has \cite{Polychronakos:2001mi}
\begin{equation}\label{eq:Zcl}
    Z_\text{cl} = \sum_{n=1}^{N-1} \sqrt{k n} |n-1)(n| \,.
\end{equation}
At large $k$ (but still small compared to $N$, see footnote \ref{largek}) the matrix degrees of freedom become semi-classical and so we can use the classical state as a starting point in this limit. In particular we can, as previously, write $Z = U Z_\text{cl} U^\dagger$ to obtain the gauge orbit of ground states. Note that large $k$ was not assumed in our previous study of the vertical cut, as the exact wavefunction for the eigenvalues of $X$ was known.

Analogously to the case of the vertical cut, all matrices can be decomposed into blocks, for example
\be\label{eq:Zblock}
Z =  \left(
\begin{array}{cc}
    Z_{II} & Z_{IO} \\
    Z_{OI} & Z_{OO}
\end{array}
\right) \,,
\ee
where $O$ is for `outside'. Inserting $\Theta_I$ into the kinetic term Lagrangian as in the discussion around (\ref{eq:Lpart}), and again setting $V=1$, one has
\begin{align}
    L_{I\,\text{kin}} & = \Tr\left[Z_{II}^{\dag} \frac{d}{dt} Z_{II} + Z_{OI}^{\dag}\frac{d}{dt} Z_{OI}\right] \,, \label{eq:LI}\\
    L_{O\,\text{kin}} & = \Tr\left[Z_{OO}^{\dag} \frac{d}{dt} Z_{OO} + Z_{IO}^{\dag}\frac{d}{dt} Z_{IO}\right] \,. \label{eq:LO}
\end{align}
Note that here $Z_{OI}^\dagger = (Z^\dagger)_{IO}$.

From (\ref{eq:Zcl}), the matrix $Z^\text{cl}_{IO}$ is seen to have only one nonzero singular value $\lambda = \sqrt{k M}$. As for the vertical cut previously, this singular value is proportional to the length of the cut (recall from (\ref{eq:wigner}) that the radius of the full droplet is proportional to $\sqrt{kN}$). Having only a single nonzero singular value is a significant simplification relative to the case of the vertical cut and removes the complications with imposing the permutation symmetry and with the orthogonality constraint between the vectors associated to distinct singular values. In particular, we can represent
\be\label{eq:ZIO}
Z_{IO} = \sqrt{k M} \xi_{I}\xi_{O}^{\dag} \,,
\ee
in terms of just two vectors. Following the same logic as for the vertical cut previously, upon quantization these vectors will obey the commutators
\be\label{eq:xiIcomm}
[\xi_{Ia}, \xi_{Ib}^\dagger] = \frac{\delta_{ab}}{k M} \,,
\ee
and vectors that are related by permutations of indices should be identified.

The matrix $Z^\text{cl}_{OI}$ however is identically zero, again from (\ref{eq:Zcl}). Thus it may appear that there are no `boundary modes' for the inner region described by (\ref{eq:LI}). However, the gauge theoretic entanglement between the inner and outer regions works differently in this case compared to the case of the vertical cut. The entanglement is determined by Gauss's law (\ref{eq:gauss}). In particular, restricting to the $V=1$ gauge orbit of the classical ground state, so that (\ref{eq:ZIO}) holds and $Z_{OI} = 0$, the Gauss law corresponding to charge inside $U(M)$ is
\begin{align}\label{eq:ingauss}
    k\, \text{Id}_M = [Z_{II},Z_{II}^{\dag}] + Z_{IO}Z_{IO}^{\dag} = [Z_{II},Z_{II}^{\dag}] + k M \xi_I \xi_I^\dagger \,. 
\end{align}
Note that $\Psi$ does not appear because when the classical ground state is expressed in a basis with $R^2$ diagonalized, then the only nonzero component of $\Psi$ is in the outer region \cite{Polychronakos:2001mi}. Equation (\ref{eq:ingauss}) is different from what occurs in the vertical cut case. In that case the Gauss law acting on each region does not involve the boundary modes, and the gauge-theoretic entanglement comes from the `off-diagonal' parts of the Gauss law. In (\ref{eq:ingauss}) we see that the boundary modes $\xi_I$ that appear in the {\it outside} dynamics, via (\ref{eq:LO}) and (\ref{eq:ZIO}), are related through Gauss's law to the inner degrees of freedom $Z_{II}$.

It follows that tracing out the inner degrees of freedom will generate a mixed state for the $\xi_I$. This state must be invariant under $U(M)$. However, the boundary cut $\xi_I$ modes are the only degrees of freedom in the outside region that are charged under $U(M)$, that acts on the interior. We must therefore write down a $U(M)$ invariant state of the $\xi_I$ oscillators that further obeys the normalization constraint $\xi_I^\dagger \cdot \xi_I = 1$. From the commutation relations (\ref{eq:xiIcomm}), this requires $kM$ quanta. As in our previous discussion of the vertical cut, the only such state is the maximally mixed state in this fixed number sector. The discussion here, however, is easier. With similar manipulations to the vertical cut case
\begin{align}
\rho_{\xi_I} & = \left(\sum_a \xi_{Ia}^\dagger \otimes \xi_{Ia} \right)^{kM} |0\rangle \otimes \langle0| \\
& = (kM)! \sum_{\sum p_a = k M} |\{p_a\}\rangle \langle \{p_a\}| \,. \label{eq:ximixed}
\end{align}
As previously, the states are labelled by all possible sets of occupation numbers $\{p_a\}_{a=1}^M$ of the $M$ oscillators such that $\sum_a p_a = k M$. The final state (\ref{eq:ximixed}) is maximally mixed on the sector with $kM$ quanta.

Also as in the vertical cut case, but now with fewer complications, permutation symmetry requires that the distinct components $\xi_{Ia}$ be treated as identical operators. We therefore project the sum in (\ref{eq:ximixed}) to a sum over unlabelled $M$-partitions of $kM$. This greatly reduces the entanglement. The entanglement entropy of (\ref{eq:ximixed}) is the logarithm of the dimension of this Hilbert space. As for the vertical cut, this is given (at large $k$) by the Hardy-Ramanujan formula. For the circular cut we have
\begin{align}
s^\text{circular}_\text{cut} & = \pi  \sqrt{\frac{2 kM}{3}} - \log (4 \sqrt{3} k M) + \cdots \\
& = \frac{(Nk)^{1/2}}{\sqrt{6}} C_o - \log (\sqrt{3} N k \, C_o^2/\pi^2)  + \cdots \,. \label{eq:C}
\end{align}
Here $C_o = 2 \pi \sqrt{M/N}$ is the circumference of the boundary between the inner and outer regions, in units where the droplet has unit radius.

The final line (\ref{eq:C}) has again recovered a leading area law. The prefactor of the area law is very similar to that obtained for the vertical cut in (\ref{eq:sbulk}), but now without the logarithmic violation of the area scaling. We saw previously that the logarithmic violation was associated to having multiple nonzero singular values, whereas the circular cut only has one. Further, it is natural to suspect that this is correlated with the fact that the vertical cut intersects the boundary of the droplet while the circular cut does not. To test this correlation, in Appendix \ref{app:square} we have studied a `square' cut that, as a function of the size of the square, is either wholly contained in the droplet or intersects the boundary. We find that precisely when the square intersects the boundary, multiple singular values appear and concomitantly a logarithm appears in the area law.

The second, additive logarithmic term in (\ref{eq:C}) is under better control than in the vertical cut, as there is no ambiguity with implementing the orthogonality between singular vectors. However there remain other potential sources of logarithmic contributions to the entanglement that we will discuss further in section \ref{sec:topological} below. Before doing so we briefly consider an illuminating setting in which precisely two singular values appear.

\subsubsection{An annular cut}
\label{sec:annulus}

Consider two concentric circular boundaries, at radii $\sqrt{kM_1}$ and $\sqrt{kM_2}$. Denote the annulus in between these boundaries by
$\Sigma$. We can generalize the computation we have just performed for the circular cut to obtain the entanglement between the annulus and the (now disconnected) complementary region $\bar \Sigma$. The gauge group in this case naturally breaks down to $U(M_1) \times U(M_2 - M_1) \times U(N - M_2)$, where we additionally have fixed the unitary transformations that mix the disconnected components of $\bar{\Sigma}$. One can immediately see from $Z_\text{cl}$ in (\ref{eq:Zcl}) that $Z_{\Sigma \bar{\Sigma}}$ has two nonzero singular values of magnitude $\lambda_1 = \sqrt{kM_1}$ and $\lambda_2 = \sqrt{kM_2}$. Thus, a nontrivial topology also leads to multiple singular values.

The additional gauge fixing we have just performed, of unitaries that act within $\bar \Sigma$ by mixing the disconnected components, is perhaps the simplest possible example of the proposal described in section \ref{sec:state} for implementing the orthogonality constraint between different singular vectors. The proposal seems especially well-motivated in this present case, where it allows independent counting problems for the two boundaries. Specifically,
the interior boundary of the annulus ($\partial_1$) will have $kM_1$ quanta distributed among $M_1$ oscillators, while exterior boundary ($\partial_2$) will have $kM_2$ quanta distributed amongst $M_2 - M_1$ oscillators. Again assuming a geometric regime ($k \ll M_i$, $\sqrt{kM_2}\log{\sqrt{kM_2}} \ll M_2 - M_1$), the Hardy-Ramanujan formula applies and gives the gauge-theoretic entanglement entropy
\begin{equation}
    \sum_i \log \dim \mathcal{H}_{\partial_i} = \pi \sqrt{\frac{2}{3}}\left(\sqrt{kM_1} + \sqrt{kM_2}\right) - \log(M_1) - \log(M_2) - 2\log(4\sqrt{3}k) \,.
\end{equation}
There is again a leading perimeter law entanglement along with a logarithmic correction.

\section{Discussion}

In this section we will discuss some open questions, further directions for research and potential connections of our work to existing results. Let us first recall the big picture motivation for our work. Semiclassical analyses of gravitating spacetime suggest that there is a universal large but finite entanglement entropy associated to spatial partitions \cite{Ryu:2006bv, Hubeny:2007xt, Faulkner:2013ana, Engelhardt:2014gca}. This leads to the question: What partition of the underlying `pre-geometric' quantum degrees of freedom do these semiclassical spatial partitions correspond to? In this work we have attempted to answer this question in a relatively simple large $N$ matrix quantum mechanics that underpins an emergent two dimensional space.

We have been able to compute the gauge-theoretic entanglement entropy by mapping the calculation onto a counting problem that is solved by the Hardy-Ramanujan formula. This is the same counting that Strominger and Vafa famously used to reproduce the entropy of certain supersymmetric black holes \cite{Strominger:1996sh}. In both cases the key degrees of freedom are the modes of a chiral boson. The need to understand the microscopic origin of the Ryu-Takayanagi entanglement entropy is an update on the quest to microscopically derive the Bekenstein-Hawking black hole entropy. It seems likely that the many insights and techniques developed for counting the entropy of black holes may also be applied to the case of entanglement entropy, extending the connection we have made in this work.

We have emphasized that the gauge-theoretic entanglement appears in addition to the collective field contribution. The collective field entanglement has previously been obtained for the matrix quantum mechanics of a single matrix that describes the emergent spatial dimension of two dimensional string theory \cite{Das:1995vj, Das:1995jw, Hartnoll:2015fca}. The ground state of that theory is closely related to the state that we have discussed here in the case that $k=1$, which can equivalently be described in terms of free fermions. The length $L$ of the bulk cut that we have considered becomes the depth of the Fermi sea in this description. In two dimensional string theory $L \sim 1/g_\text{s}$, the inverse string coupling (e.g.~\cite{Jevicki:1993qn}). Our result (\ref{eq:sbulk}) therefore suggests that in two dimensional string theory there is a gauge-theoretic entanglement entropy $s_\text{gauge} \sim 1/g_\text{s}$ that should be added to the previously obtained collective field entropy $s_\text{col} \sim \log (1/g_\text{s})$.

Perhaps the most central of our results is that the gauge-theoretic entanglement in our matrix quantum mechanics is controlled by a small number of large singular values of the `off-diagonal' matrix elements. We found that these singular values are given by a length in the emergent geometry. This provides a rather direct connection between entanglement and geometry that may be useful to characterize emergent geometries more generally. It will be important to understand whether this connection can be generalized to higher dimensional emergent geometries and also to compressible models with bulk excitations. For the latter case, the fuzzy sphere ground state of the `mini-BMN' model is a natural starting point \cite{Han:2019wue}.

We end with an extended discussion of two open issues arising in our work.

\subsection{Comments on gauge symmetry and the Moyal map}
\label{sec:moyal}

The gauge symmetry of our matrix quantum mechanics is $U(N)$.
We have seen how the gauge-theoretic entanglement of the model is closely connected to the gauge-theoretic entanglement of the local $U(1)$ gauge symmetry in Chern-Simons theory. A direct connection between these two groups can be made via a Moyal-type map from the matrices to the noncommutative plane. Consider a perturbation of the ground state $X = X_\text{cl} + A_y$ and perform an infinitesimal unitary transformation generated by $T$ on $X$. The leading order change of $X$ is $i[T,X_\text{cl}]$ which under the Moyal map becomes $\pa_y T$. Under the Moyal map this transformation therefore corresponds to the local $U(1)$ gauge transformation $A_y \to A_y + \pa_y T$. More precisely the correspondence goes via a non-commutative $U(1)$ gauge symmetry \cite{Susskind:2001fb, FRADKIN2002483}. At next order the transformation is furthermore seen to generate an area preserving diffeomorphism on $A_y$ --- see \cite{Han:2019wue} for a recent related discussion for the case of the fuzzy sphere.

The quantum Hall matrix model we have discussed can be generalized to describe non-Abelian quantum Hall states \cite{Dorey:2016mxm, Dorey:2016hoj}. The corresponding Chern-Simons theories have larger local gauge groups. It will be instructive to understand how the gauge-theoretic entanglement is encoded by the matrix degrees of freedom in those cases.

The Moyal map also gives insight into 
the geometrical meaning of the off-diagonal $Y_{LR}$ matrices that have played a central role in our discussion. Consider the Moyal map
\be\label{eq:moyal}
[Y_\text{cl},\Theta_\text{cl}] = [Y_\text{cl},\theta(X_\text{cl} - x_o \text{Id})] \to i \pa_x \theta(x-x_o) = i \delta(x-x_o) \,.
\ee
Here $\Theta_\text{cl}$ is the projector defined in (\ref{eq:Theta}), without the factors of $U$, and $\theta$ is the heaviside step function. We see in (\ref{eq:moyal}) that
the matrix $[Y_\text{cl},\Theta_\text{cl}]$ is naturally associated to functions that are localized close to $x=x_o$.
However, this matrix can be written explicitly as
\be
[Y_\text{cl},\Theta_\text{cl}] = \left(
\begin{array}{cc}
    0 & - Y^\text{cl}_{LR} \\
    Y^\text{cl}_{RL} & 0
\end{array}
\right) \,.
\ee
Consider for simplicity that $Y^\text{cl}_{LR}$ has a single nonzero singular value $\lambda_\text{cl}$. The singular value can be extracted by squaring and taking the trace: $\tr \left( Y^\text{cl}_{LR} Y^{\text{cl} \, \dagger}_{LR}\right) = \lambda^2_\text{cl}$. However, under the Moyal map, the trace of the matrix corresponds to the integral of the function. Squaring the right hand side of (\ref{eq:moyal}) gives $\delta(x-x_o) \star \delta (x-x_o) \propto L \delta(x-x_o)$. See e.g. \cite{Minwalla:1999px} for discussions of function multiplication under the Moyal map; we have naively cut off an IR divergence in the $y$ direction by the length across the droplet at $x=x_o$. Integrating this single delta function we obtain $\lambda_\text{cl}^2 \propto L^2$. That is to say, 
$\lambda_\text{cl}$ is the length of the cut, precisely the result we found previously from an explicit computation! Our statements here are not fully rigorous because the Moyal map is complicated by geometric boundaries. These complications may be related to the appearance of multiple singular values when the cut intersects the boundary. Nonetheless, the connection we have outlined indicates that the geometrical interpretation of the singular values of the off-diagonal matrix may be quite robust.
We hope to make this general connection more precise in future work.

\begin{figure}[h!]
    \centering
    \includegraphics[width=0.45\textwidth]{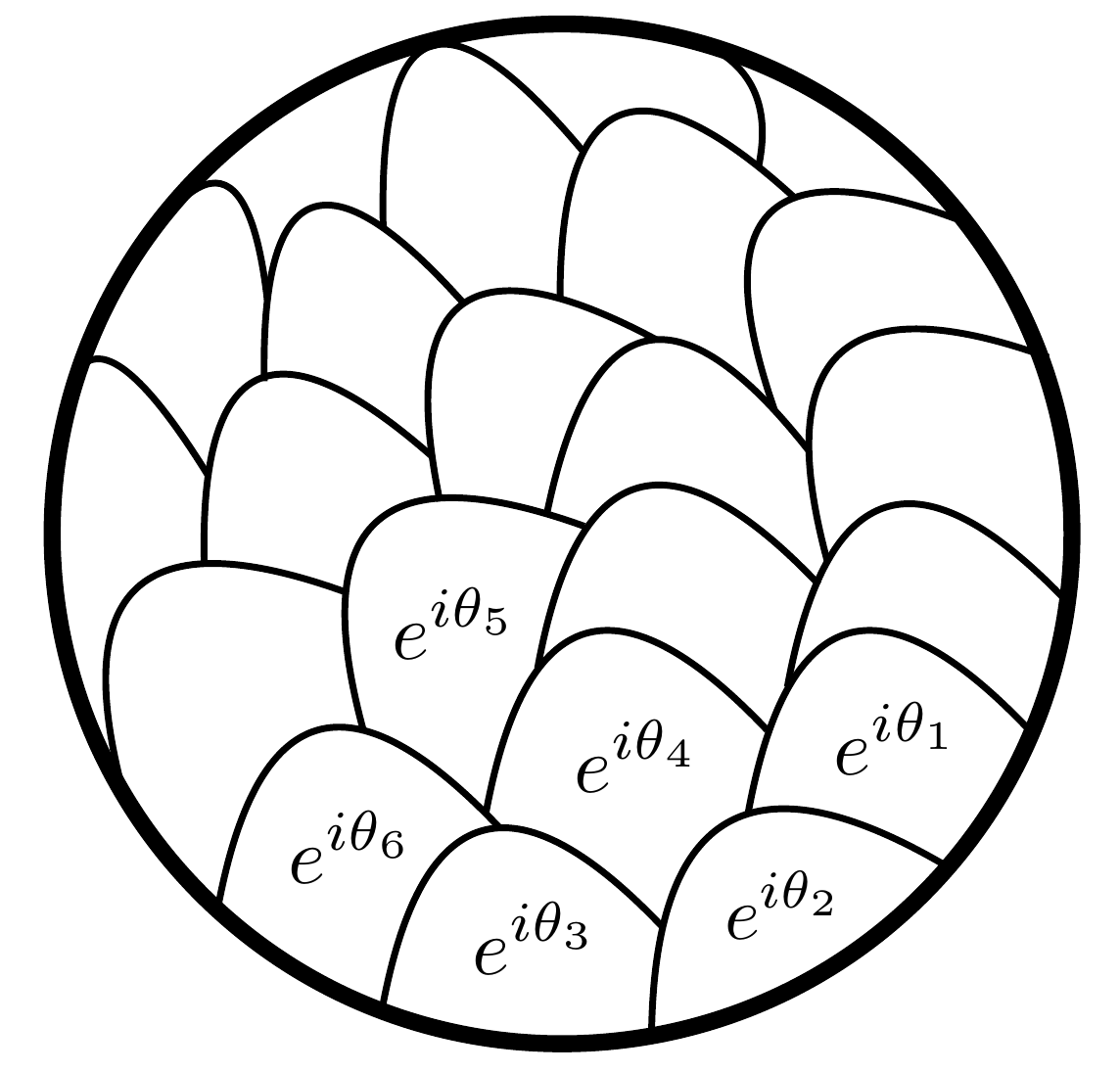}
    \caption{The Moyal map associates a unitary matrix to a collection of $N$ phases, several of which are shown, each within non-overlapping regions of unit area.}
    \label{fig:unit}
\end{figure}

We can now consider $Y_{LR} = U_L Y^\text{cl}_{LR} U_R^\dagger$. Unitary matrices can be written as a sum of projectors weighted by phases: $U = \sum_s e^{i \theta_s} v_s v_s^\dagger$. Under the Moyal map each projector $P_s \equiv v_s v_s^\dagger$ will map onto a function. Because $\sum_s P_s = \text{Id}$, $\text{tr} P_s = 1$ and $\text{tr} (P_s P_t) = \delta_{st}$, these projectors are naturally associated to non-overlapping regions of the droplet, each with unit area. This leads to the picture of a unitary matrix shown in Fig. \ref{fig:unit}.
Now, $U_L$ and $U_R$ correspond to unitaries that are only nontrivial on a single side of the cut. The matrix $Y_{LR} = U_L Y^\text{cl}_{LR} U_R^\dagger$ is therefore seen to be associated to a sum of differences in phases (due to $U_L$ and $U_R^\dagger$) of the regions that lie adjacent to the cut, so the corresponding projectors are nonzero when integrated against the delta function (\ref{eq:moyal}). Schematically, $Y_{LR} \to  e^{i [\theta_L(x_o,y) - \theta_R(x_o,y)]} \delta(x-x_o)$.
This difference in phases along the entanglement cut is the data expected for the boundary chiral mode in Chern-Simons theory.

Finally, the discussion above enables us to make some comments about how the $U(N)$ gauge symmetry encodes area-preserving diffeomorphisms as well as the local $U(1)$ gauge symmetry.
Returning to our starting point in section \ref{sec:intro}, pick a matrix $F = f(X,Y)$. This matrix can be diagonalized as $F = \sum_s F_s v_s v_s^\dagger$. The unit area regions corresponding to the projectors $v_s v_s^\dagger$ will now be (fuzzy) level sets of $f(X,Y)$. For example, they will be vertical stripes when $F = X$ and concentric rings when $F = R^2$. Upon acting with a general unitary, so that $F' = U F U^\dagger$, $F'$ will not be diagonal in the same basis. Diagonalizing $F' = \sum_s F'_s v'_s v_s'{}^{\dagger}$ leads to a new set of regions corresponding to the projectors $v'_s v_s'{}^{\dagger}$. That is, the unitary matrix shuffles the unit area regions around the droplet.

The $U_L$ and $U_R$ unitaries act on complementary eigenspaces of $F$. Therefore, they only shuffle around regions that lie on a given side of the cut. The $V$ unitaries, in contrast, can lead to regions that intersect the cut. One may suspect that such regions, that extend across the cut, can act as a buffer that reduces the entanglement between purely left and right degrees of freedom.  We hope to explore this possibility in the future. It should be clear, also, that an understanding of the matrix entanglement from the perspective of diffeomorphisms (rather than the local U(1)), will be essential in theories with an emergent gravitating spacetime.

\subsection{Comments on the logarithmic term}
\label{sec:topological}

The subleading logarithmic terms that have appeared in the results above, such as (\ref{eq:C}), are reminiscent of the topological entanglement entropy \cite{PhysRevLett.96.110404, PhysRevLett.96.110405}. For a circular cut in Chern-Simons theory this would be $- \frac{1}{2} \log k$. We now list several issues that will need to be clarified in order to make a precise connection between the matrix entanglement and the topological entanglement.

Firstly, it is not obvious a priori that topological entanglement should be present in the matrix model, because the microscopic completion does not have an in-built two dimensional spatial locality. Space itself emerges with the ground state.

Secondly, the counting problem for the Chern-Simons entanglement includes a contribution from winding (or `zero') modes around the circular cut. We did not write these down in (\ref{eq:sCS}), yet this contribution is essential to obtain the subleading $-\frac{1}{2} \log k$ term in Chern-Simons theory \cite{Belin:2019mlt}. There does not appear to be an analogue of these terms in the matrix model counting problem as we have set it up here.

Thirdly, there are additional potential sources of logarithmic entanglement in the circular cut. Away from the classical limit, the singular value $\lambda$ will fluctuate (the classically vanishing singular values will also fluctuate but these are unimportant for the entanglement). This leads to an additional term in the entanglement entropy analogous to the first term in (\ref{eq:ptot1}) above, namely $- \sum_\lambda p(\lambda) \log p(\lambda)$, where
$p(\lambda)$ is the probability distribution of the singular values $\lambda$. We can estimate this contribution by taking the fluctuations of $\lambda$ to be Gaussian with variance $\Delta \lambda^2$. In that case one immediately finds an additive logarithmic correction to the entropy
\be
\Delta_\lambda s^\text{circular}_\text{cut} = \log (\Delta \lambda) + \cdots \,.
\ee
There is an analogous additional contribution if the entanglement cut is defined geometrically in terms of the radius $r_o$ rather than the number of eigenvalues $M$. As we noted for the vertical cut in footnote \ref{eq:Mfluc}, a geometric cutoff will lead to an uncertainty $\Delta M$ in the eigenvalues on a given side of the cut. These lead to an additional entanglement
\be
\Delta_M s^\text{circular}_\text{cut} = \log (\Delta M) + \cdots \,.
\ee
Further to this point, it may be the case that the microscopic partition that corresponds to a geometric field theoretic partition should be smeared over a lengthscale set by the eigenvalue granularity. This will lead to an additional uncertainty $\Delta M$.

We hope to address these issues in a future more detailed study of the circular cut.

\section*{Acknowledgments}

We are especially grateful to Onkar Parrikar and Xizhi Han for extensive discussions and collaboration at the early stages of this work. We thank Ronak Soni for helpful comments on the first version of this paper. This work was partially supported by DOE award DE-SC0018134, by Simons Investigator award \#620869 and by STFC consolidated grant ST/T000694/1. S.A.H.~acknowledges the hospitality of the Max Planck Institute CPfS while part of this work was carried out. A.F. is supported by the National Science Foundation Graduate Research Fellowship under Grant No. DGE-1656518.

\appendix

\section{Constraints from the measure}
\label{sec:appA}

In this appendix we perform a large $N$ saddle point evaluation of the normalization condition (\ref{eq:norm2}) and verify that the measure term imposes normalization and positivity of the collective field. We simultaneously do the saddle point analysis for the collective field $n$ and the field $\lambda$ appearing in the measure (\ref{eq:J}). Using the wavefunction (\ref{eq:psin}), the saddle point equations are
\begin{align}
2 k \int dy n(y) \log|x - y| & = x^2 - i \lambda(x) \,, \label{eq:first} \\
n(x) & =  \frac{N e^{-i \lambda(x)}}{\int dy e^{-i \lambda(y)}} \,.
\end{align}
The second equation here implies that
\be
\int n(x) dx = N \,.
\ee
Thus normalization of the collective field is imposed at this order. In fact, we can solve this second equation by setting
\be\label{eq:lambda}
i \lambda(x) = -\log \frac{n(x)}{\mu} \,,
\ee
here $\mu$ is a constant that will be determined by normalization. Now plugging (\ref{eq:lambda}) into (\ref{eq:first}) and differentiating with respect to $x$ we obtain
\be
k \fint \frac{dy \, n(y)}{x-y} = x + \frac{1}{2} \frac{n'(x)}{n(x)} \,.
\ee
This equation can be found in e.g. \cite{Andric_1988}. The final term smooths out the eigenvalue distribution so that it doesn't have a sharp edge, but rather an exponential tail. However, this term is subleading at large $N$, that is to say, the exponential decay away from the droplet is very fast. To see this note that the large $N$ scaling of the saddle point is
\be
n(x) = \sqrt{N} \hat n(\hat x) \,, \qquad x = \sqrt{N} \hat x \,.
\ee
With this scaling the final term is suppressed by a factor of $1/N$ relative to the other two terms. Dropping this term, the remaining terms are then solved by the Wigner semi-circle
\be\label{eq:wigner2}
n_o(x) = \frac{2 N}{\pi R^2} \sqrt{R^2 - x^2} \,, \qquad R^2 = 2N k \,.
\ee
We have normalized the solution. The other field is then
\be
\lambda_o(x) = i \log n_o(x) \,. 
\ee

To go to next order in the saddle point (large $N$) expansion we can set
\be
n(x) = n_o(x) + \delta n(x) \,, \qquad \lambda(x) = \lambda_o(x) + \delta \lambda(x) \,.
\ee
The normalization constraint is then
\be
\int {\mathcal D \delta n} {\mathcal D \delta \lambda} \, e^{F_2[\delta n, \delta \lambda]} = 1 \,,
\ee
with the exponent to quadratic order being
\begin{align}
F_2[\delta n, \delta \lambda] = k & \int dx_1 dx_2 \delta n(x_1) \delta n(x_2) \log |x_1 - x_2| + i \int dx \delta n(x) \delta \lambda(x) \nonumber \\
& - \frac{1}{2} \int dx n_o(x) \left[\delta\lambda(x) - \frac{1}{N} \int dy n_o(y) \delta \lambda(y) \right]^2 \,.
\end{align}
Performing the $\delta \lambda$ integral gives
\be\label{eq:psi33}
\int {\mathcal D \delta n} \, \delta\left(\textstyle \int dx \delta n(x)\right) e^{k \int dx_1 dx_2 \delta n(x_1) \delta n(x_2) \log |x_1 - x_2| - \frac{1}{2} \int dx [\delta n(x)^2/n_o(x)]} = 1\,.
\ee
The last term in the exponent clearly forces $\delta n(x)$ to be zero outside of the support of $n_o(x)$. The overall delta function forces the integral of $\delta n$ to vanish, thus preserving normalization. This delta function arises from doing the integral over the zero mode $\int n_o(y) \delta \lambda(y) dy$.

The last term in the exponent of (\ref{eq:psi33}) is subleading at large $N$ inside the support of $n_o(x)$. Therefore we may drop it after imposing the boundary condition that $\delta n(x)$ vanish outside the support $x \in [-R,R]$.

\section{Eigenvalue to collective field entanglement}
\label{sec:appB}

Starting with the quantum mechanical description, upon tracing out the region $x < x_o$ the reduced density matrix $\rho_\text{red}$ for the eigenvalues is split into sectors in which there are $s$ eigenvalues in the region $x < x_o$:
\be
\rho_\text{red} = \sum_{s=0}^N \int d^sz d^{N-s}x d^{N-s}y \, \overline \psi(z,x) \psi(z,y) |x\rangle \langle y| \,.
\ee
Here the eigenvalue wavefunction $\psi$ has been written in terms of the eigenvalues that lie to the left and right of $x_o$. Recall that the eigenvalues have been ordered. The $n$th trace of the reduced density matrix is then
\be
\tr \rho_\text{red}^n = \sum_{s=0}^N \int \prod_{i=1}^n d^s z_i d^{N-s} x_i \overline \psi(z_i,x_i) \psi(z_i,x_{i+1}) \,.
\ee
Here $x_{n+1} \equiv x_1$. This expression can now be written in terms of collective fields
\be\label{eq:trace}
\tr \rho_\text{red}^n = \sum_{s=0}^N \int \prod_{i=1}^n {\mathcal D} n^<_i {\mathcal D} n^>_i J_s\left[n^<_i\right] J_{N-s}\left[n^>_i\right] \overline \psi\left[n^<_i,n^>_i\right] \psi\left[n^<_i,n^>_{i+1}\right] \,.
\ee
Note that $s$ only appears in the measure factors. The collective fields $n_i^<(x) = \sum_{a=1}^s \delta(x-z_{ia})$ and $n_i^>(x) = \sum_{a=s+1}^N \delta(x - x_{ia})$ are supported on $x < x_o$ and $x > x_o$, respectively.

We know from Appendix \ref{sec:appA} that in the large $N$ limit the role of the measure factor is to impose the normalization and positivity of the collective field. The normalization constraints are now
\be\label{eq:newnorm}
\int_{-R}^{x_o} dx n^<_i(x) = s \,, \qquad \int_{x_o}^R dx n^>_i(x) = N - s \,.
\ee
The sum over $s$ can be approximated as an integral at large $N$. Thus we can write, setting $n_i = n_i^< + n_i^>$ and $\widetilde n_{i} = n_i^< + n_{i+1}^>$,
\be\label{eq:trace2}
\tr \rho_\text{red}^n = \int \prod_{i=1}^n {\mathcal D} n_i \mu\left[n_i, \widetilde n_i \right] \overline \psi\left[n_i\right] \psi\left[\widetilde n_{i}\right] \,.
\ee
Here the measure
\bea
\mu\left[n_i, \widetilde n_i \right] & = & \int ds \prod_{i=1}^n \delta\left(\int dx n^>_i(x) - (N-s) \right) \delta\left(\int dx n^<_i(x) - s \right) \,. \label{eq:cons}
\eea
Thus a constraint remains on the collective field partition. Not only is the total normalization fixed, but also the normalization of the region that is traced over.

To see the effect of the constraints in (\ref{eq:cons}) in computing the traces (\ref{eq:trace2}), write the wavefunction $\psi[\delta n]$ as (\ref{eq:psi4}) in the main text and split $\phi_i = \phi_i^< + \phi_i^>$ into its components with support on $x < x_o$ and $x > x_o$. Given $x_o$, the value of $s = s_o$ on the saddle point is fixed and therefore the constraints on the fluctuations becomes
\bea
\mu\left[\delta n_i, \delta \widetilde n_i \right] & = & \int d \delta s \prod_{i=1}^n \delta\left(\int dx \delta n^>_i(x) + \delta s \right) \delta\left(\int dx \delta n^<_i(x) - \delta s \right) \,. \label{eq:cons2}
\eea
That is, while the fluctuation does not change the total number of eigenvalues, it can re-distribute the eigenvalues between the two regions. Performing the integrals over the $\delta n_i$ one obtains
\begin{align}
\tr \rho_\text{red}^n & = \int \prod_{i=1}^n {\mathcal D} \mu^<_i {\mathcal D} \mu^>_i {\mathcal D} \phi_i {\mathcal D} \widetilde \phi_i \overline \Phi[\phi_i] \Phi[\widetilde \phi_i] \, \delta\left[ \phi_i^< - \widetilde \phi_i^< + \mu^<_i \right] \delta\left[\phi_i^> -  \widetilde \phi_{i-1}^> + \mu^>_i \right] \nonumber \\
& \qquad \qquad \times \delta\Big(\sum_i \left[\mu_i^< - \mu_i^>\right] \Big)  \,.\label{eq:eeb}
\end{align}
Here the $\mu^<_i$ and $\mu^>_i$ are constant Lagrange multipliers that impose the constraints in (\ref{eq:cons2}). This is almost the entanglement of the state $\Phi$ under a spatial partition. The only complication is that the constant modes of $\phi$ and $\widetilde \phi$ in each region are not identified, but are rather shifted by the $\mu$ variables.

The expression (\ref{eq:eeb}) can be re-written as follows.
Firstly, we note that because the total fluctuation preserves the number of eigenvalues, so that $\int dx \delta n = 0$, then the field $\phi$ introduced in (\ref{eq:n1a}) must not in fact include an overall constant zero mode (as the integral over this mode would not converge). To obtain the usual entanglement of a boson, we need this zero mode to be included in the traces in (\ref{eq:eeb}). We will now see that the $\mu_i$ variables in (\ref{eq:eeb}) are naturally re-interpreted as the missing zero modes. Specifically, consider a linear change of variables
\begin{align}
\mu_i^< = \phi_{oi} - \widetilde \phi_{oi} \,, \qquad
\mu_i^> = \phi_{oi} - \widetilde \phi_{o{i-1}} + \mu_\perp \,.\label{eq:ch}
\end{align}
To ensure that the number of variables matches we can fix, for example, $\widetilde \phi_{on} = c$. This is because there is a redundancy in which all the $\phi_o$ and $\widetilde \phi_o$ are shifted by the same constant. The final delta function in (\ref{eq:eeb}) becomes simply
\be
\delta\Big(\sum_i \left[\mu_i^< - \mu_i^>\right] \Big) = \delta \left(\mu_\perp \right) \,. 
\ee
Now define the new fields
\be
\phi_i^F = \phi_i + \phi_{oi} \,, \qquad \widetilde \phi_i^F = \widetilde \phi_i + \widetilde \phi_{oi} \,.
\ee
The trace becomes
\be
\tr \rho_\text{red}^n = \sqrt{n} \int \prod_{i=1}^n {\mathcal D} \phi^F_i {\mathcal D} \widetilde \phi^F_i \overline \Phi[\phi^F_i] \Phi[\widetilde \phi^F_i] \, \delta\left[ \phi_i^{F<} - \widetilde \phi_i^{F<} \right] \delta\left[\phi_i^{F>} -  \widetilde \phi_{i-1}^{F>} \right] \delta\left(\textstyle \frac{1}{\pi} \int d\theta \widetilde \phi^F_{n} - c\right) \,. \label{eq:withc}
\ee
Here we used the fact that $\Phi[\phi]$ does not depend on the constant mode in $\phi$. For this reason, (\ref{eq:withc}) does not depend on the value of $c$. The factor of $\sqrt{n}$ comes from the measure due to the change of variables (\ref{eq:ch}). Integrating over $c$ we obtain
\be
\tr \rho_\text{red}^n = \frac{\sqrt{n}}{\text{vol}\left(\textstyle \frac{1}{\pi} \int d\theta \widetilde \phi^F_{n} \right)} \int \prod_{i=1}^n {\mathcal D} \phi^F_i {\mathcal D} \widetilde \phi^F_i \overline \Phi[\phi^F_i] \Phi[\widetilde \phi^F_i] \, \delta\left[ \phi_i^{F<} - \widetilde \phi_i^{F<} \right] \delta\left[\phi_i^{F>} -  \widetilde \phi_{i-1}^{F>} \right]  \,. \label{eq:trfinal}
\ee

\section{Replica symmetry}
\label{sec:appC}

Using the wavefunction (\ref{eq:psin}), the exponent in the R\'enyi entropy (\ref{eq:trace3})
is
\begin{align}
\sum_i & \left(  S[n_i] + S[\widetilde n_i] \right) = \\
& \sum_i \left[ k \int^{x_o} dx_1 \int^{x_o} dx_2 n_i^<(x_1) n_i^<(x_2) \log |x_1 - x_2| - \int^{x_o} dx n_i^<(x) x^2 \right. \nonumber \\
& + k \int_{x_o} dx_1 \int_{x_o} dx_2 n_i^>(x_1) n_i^>(x_2) \log |x_1 - x_2| - \int_{x_o} dx n_i^>(x) x^2 \nonumber \\
& + k \int^{x_o} dx_1 \int_{x_o} dx_2 n_i^<(x_1) n_i^>(x_2) \log |x_1 - x_2| \nonumber \\
& \left. + k \int^{x_o} dx_1 \int_{x_o} dx_2 n_i^<(x_1) n_{i+1}^>(x_2) \log |x_1 - x_2|  \right]\,. \nonumber
\end{align}
This can be rewritten as follows, where we re-introduce $n_i = n_i^< + n_i^>$, with each term supported on one side of the cut:
\begin{align}
\sum_i & \left(  S[n_i] + S[\widetilde n_i] \right) = \label{eq:SS} \\
& \sum_i \left[ k \int dx_1 dx_2 n_i(x_1) n_i(x_2) \log |x_1 - x_2| - \int dx n_i(x) x^2 \right. \nonumber \\
& \left. + k \int dx_1 dx_2 n_i^<(x_1) \left( n_{i+1}^>(x_2) - n_{i}^>(x_2) \right) \log |x_1 - x_2|  \right]\,. \nonumber
\end{align}
Now perform the shift
\be
n_i(x) = n_o(x) + \varphi_i(x) \,,
\ee
with $n_o$ the (replica-symmetric) Wigner semi-circle (\ref{eq:wigner}). This extremizes the first line in (\ref{eq:SS}). Furthermore, the first order shift of the second line in (\ref{eq:SS}) is proportional to $\sum_i (\varphi_{i+1} - \varphi_i) = 0$.
It follows that the replica-symmetric Wigner semi-circle is therefore a stationary point for the computation of R\'enyi entropies.

More explicitly, the exponent in terms of the shifts is, dropping an overall constant term due to $n_o$ but otherwise with no approximation,
\begin{align}
\sum_i & \left(  S[n_i] + S[\widetilde n_i] \right) =  \\
& k \sum_i \int dx_1 dx_2 \left[ \varphi_i(x_1) \varphi_i(x_2) + \varphi_i^<(x_1) \left( \varphi_{i+1}^>(x_2) - \varphi_{i}^>(x_2) \right)\right] \log |x_1 - x_2|  \,. \nonumber
\end{align}
This quadratic function can now be shown to be negative. We can write it as
\begin{align}
\sum_i & \left(  S[n_i] + S[\widetilde n_i] \right) =  \\
& k \sum_i \int dx_1 dx_2  \left(
\begin{array}{c}
\varphi^<_i(x_1) \\ \varphi^>_i(x_1) \\ \varphi^>_{i+1}(x_1)
\end{array}
\right)^T
\left(
\begin{array}{ccc}
1 & \half & \half \\
\half & \half & 0 \\
\half & 0 & \half
\end{array}
\right)
\left(
\begin{array}{c}
\varphi^<_i(x_2) \\ \varphi^>_i(x_2) \\ \varphi^>_{i+1}(x_2)
\end{array}
\right)
\log |x_1 - x_2|  \,. \nonumber
\end{align}
The symmetric matrix appearing in this expression has eigenvalues $3/2$, $1/2$ and $0$. It is therefore positive. Furthermore, the functional product $\int dx dy f(x) f(y) \log|x-y|$ is negative for any $f$. This can be seen, for example, by Fourier transforming. It follows that replica symmetry breaking cannot increase the value of the exponent from the replica-symmetric Wigner semi-circle stationary point.

The zero mode corresponds to functions of the form $\varphi_i^<(x) = - \alpha(x) \,, \quad \varphi_i^>(x) = \alpha(x)$, for any $\alpha(x)$. This replica-symmetric deformation is not compatible with $\varphi_i^<$ being supported on one side of the cut and $\varphi_i^>$ being supported on the other side. Therefore, there are no zero modes and the replica-symmetric Wigner semi-circle is the absolute maximum.

\section{Numerical results on the singular values}
\label{sec:num}

At large $k$ the singular values of $Y_{RL}^\text{cl}$ can be obtained from the classical matrix ground states for $X$ and $Y$. In the basis in which the radius matix $R$ is diagonal --- this is the same basis that we use in section \ref{sec:circular} in the main text --- the classical ground state has \cite{Polychronakos:2001mi}
\begin{align}\label{eq:XYcl}
    X_\text{cl} & = \sqrt{\frac{k}{2}}\sum_{n=1}^{N-1} \sqrt{n} \Big[|n-1)(n| + |n)(n-1|\Big] \,, \\
    Y_\text{cl} & = - i\sqrt{\frac{k}{2}}\sum_{n=1}^{N-1} \sqrt{n} \Big[|n-1)(n| - |n)(n-1| \Big] \,.
\end{align}
The matrix $Y_{RL}^\text{cl}$ can then be obtained by diagonalizing $X_\text{cl}$ above, writing $Y_\text{cl}$ in the basis in which $X_\text{cl}$ is diagonalized, and then explicitly computing $(1-\Theta)Y_\text{cl}\Theta$. The $\Theta$ matrices are diagonal is this basis. One can then numerically determine, for some given $N$, the singular values of $Y_{RL}^\text{cl}$.

\begin{figure}[h!]
    \centering
    \includegraphics[width=0.7\textwidth]{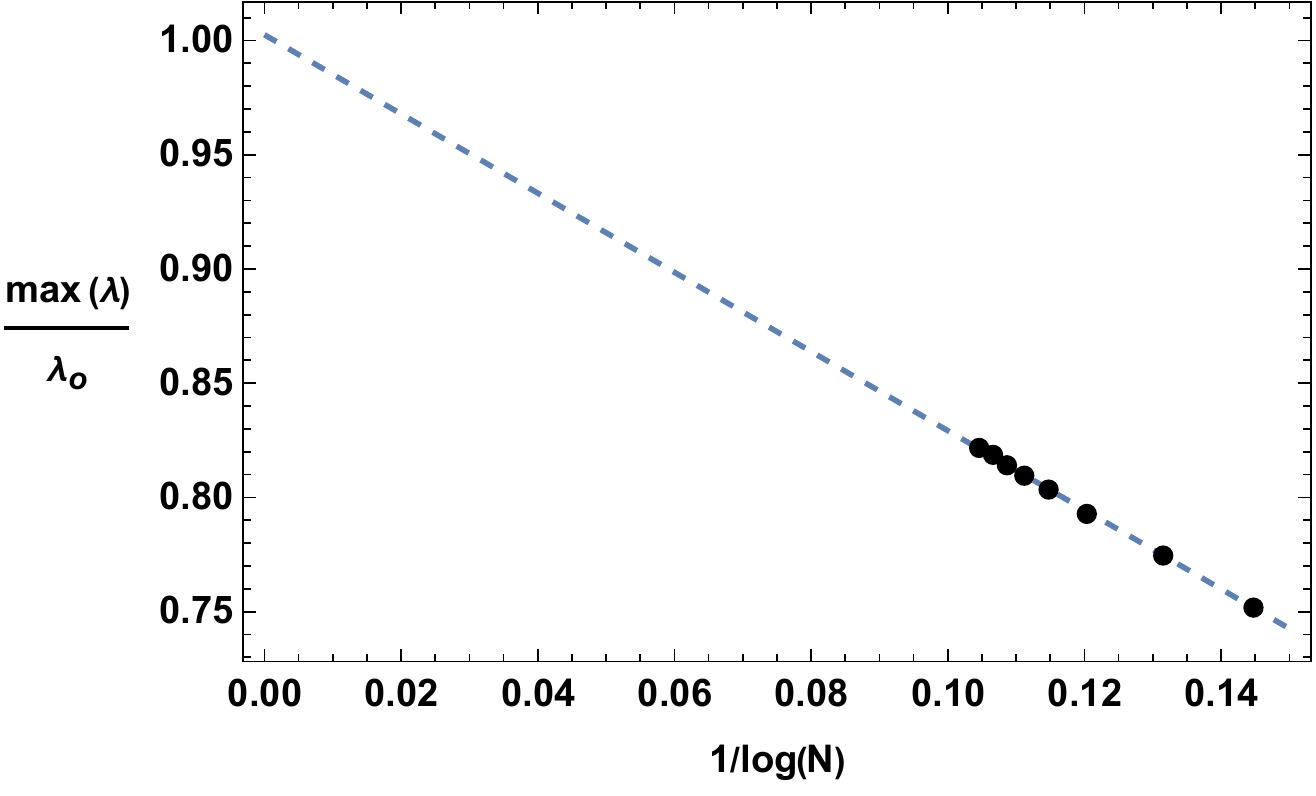}
    \caption{Black dots are numerical results for the maximal singular value of $Y_{RL}^\text{cl}$ for $N=\{1,2,4,6,8,10,12,14\}\times 10^4$. The maximal singular value is divided by the analytical prediction $\lambda_o$ given in (\ref{eq:lam0}). A linear fit of the data to $\max(\l)/\lambda_o = A + B/\log(N)$ yields the dashed line shown, consistent with convergence to the predicted value as $N \to \infty$. The results for the ratio $\max(\l)/\lambda_o$ do not depend strongly on the location $x_o$ of the cut.}
    \label{fig:logplot}
\end{figure}

Figure \ref{fig:logplot} shows the maximal singular value of $Y_{RL}^\text{cl}$ obtained numerically, as just described, as a function of $N$. The numerical value converges at large $N$ to the analytical value $\lambda_o$ obtained in (\ref{eq:lam0}). The convergence, however, is seen to be logarithmically slow as $N \to \infty$. 

\begin{figure}[h!]
    \centering
    \includegraphics[width=0.7\textwidth]{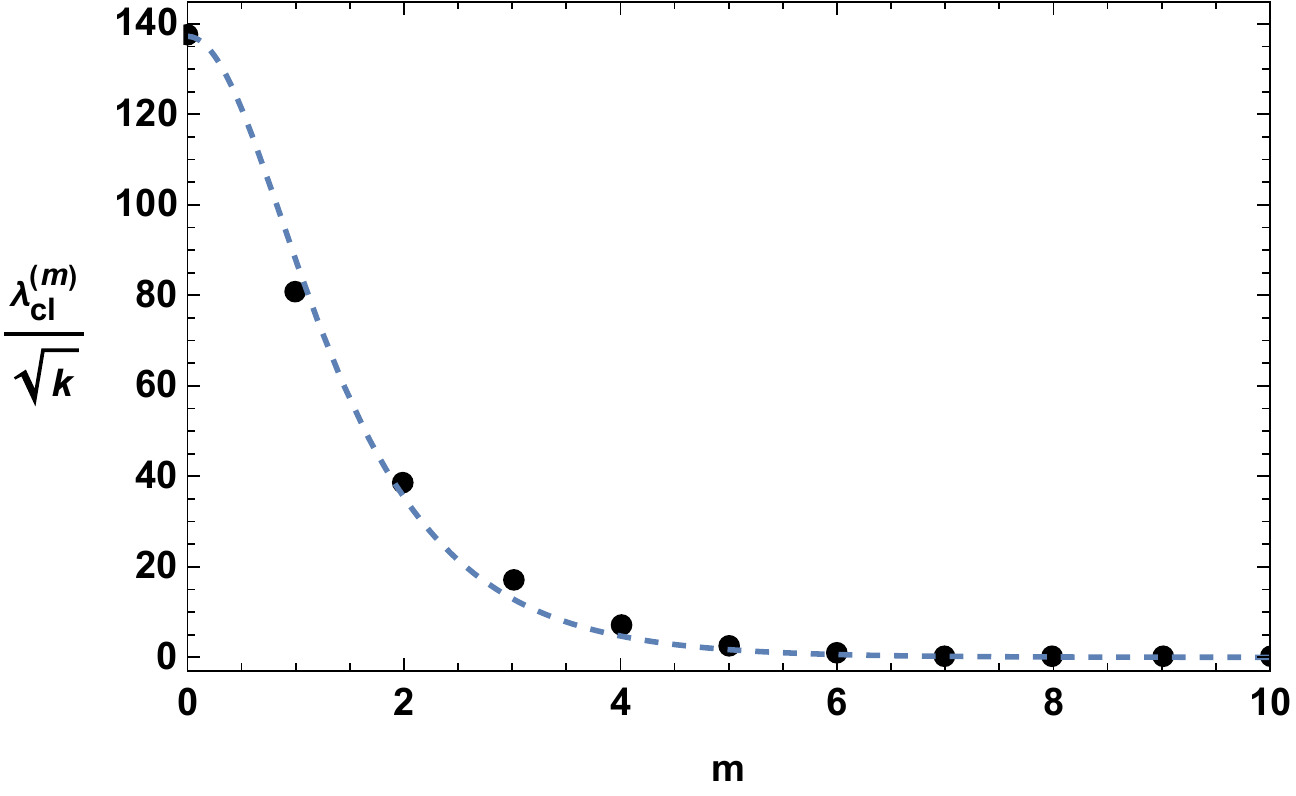}
    \caption{Black dots are singular values $\lambda^{(m)}_\text{cl}$ obtained numerically with $N=14000$ and with the cut at $x_o = 0$. The dashed line is a one-parameter fit of the data to the analytic form of (\ref{eq:lamcl}): $\lambda^{(m)}_\text{cl} = 0.82 \lambda_o \sech \frac{\pi^2 m}{\log (N/C)}$. The rescaling by 0.82 is set by figure \ref{fig:logplot} while the best fit has $C=0.85$.}
    \label{fig:sinhplot}
\end{figure}

Fixing the value of $N=14000$, figure \ref{fig:sinhplot} now shows the numerically obtained singular values $\lambda_\text{cl}^{(m)}$. The result is seen to agree well --- after an overall rescaling by a factor of $\max(\l)/\lambda_o \approx 0.82$ from figure \ref{fig:logplot} --- with the analytic expression (\ref{eq:lamcl}). In particular, only about 6 of the 14000 singular values are seen to be non-negligible.

\section{A square cut}
\label{app:square}

\begin{figure}[h!]
    \centering
    \includegraphics[width=0.45\textwidth]{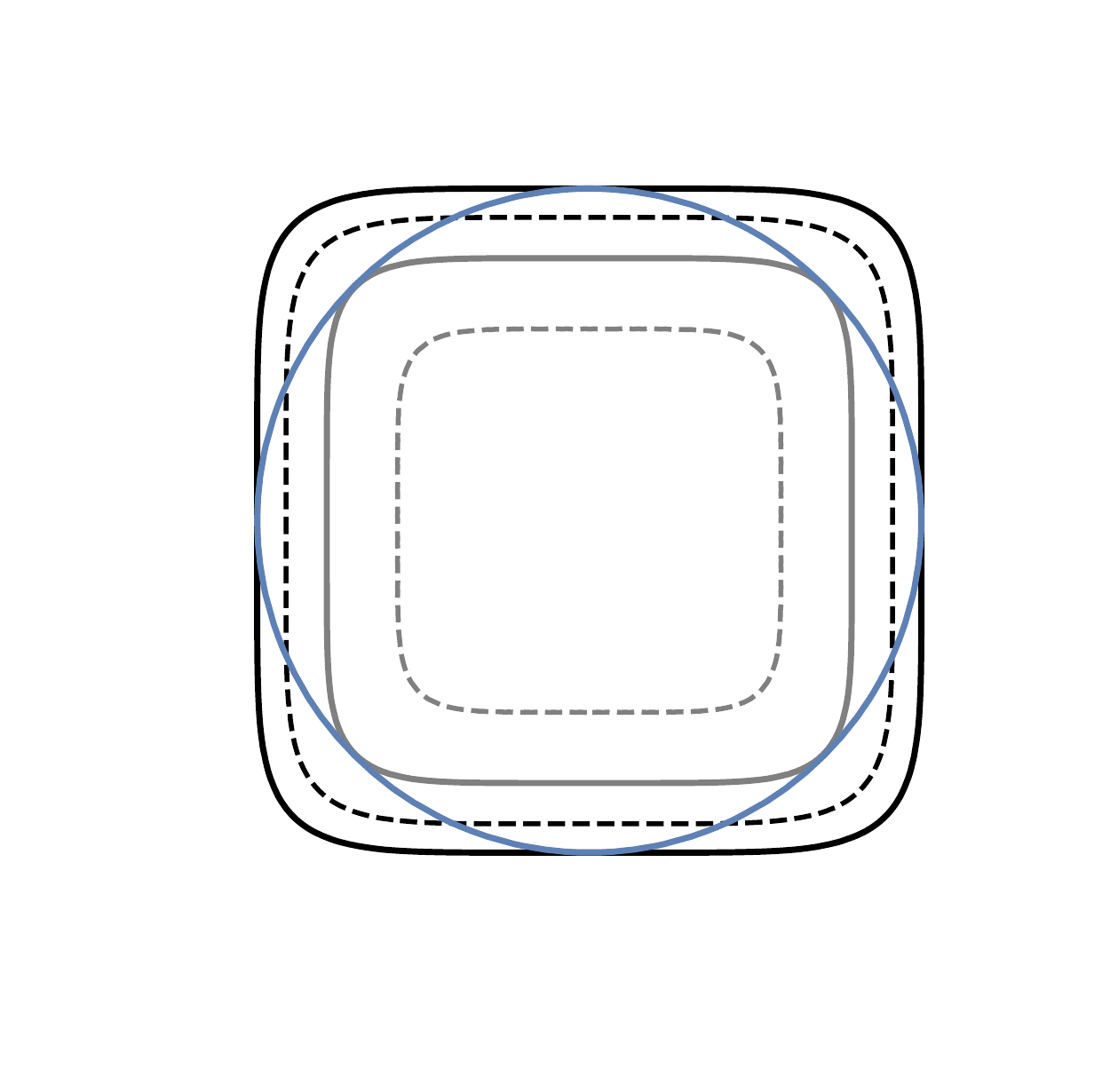}
    \caption{Illustrative examples of the square-like entanglement cut defined by $x^6 + y^6 = c^6$. The dashed gray line is fully contained within the droplet boundary (shown in blue), the solid gray line represents the transition point to a cut that intersects the boundary, the black dashed line is a typical such cut, and the solid black line is at the point where the cut leaves the disk entirely.}
    \label{fig:scAll}
\end{figure}

To explicitly show that the multiplicative logarithmic violation of the area law is associated to cuts that intersect the boundary of the quantum Hall droplet, in this appendix we consider a further square-like cut. These cuts will be defined by the level sets of $x^6 + y^6 = c^6$, as shown in Fig. \ref{fig:scAll}. At small $c$ the cuts are contained within the droplet, but beyond a critical $c$ they intersect the boundary.

To partition the geometry and extract the off-diagonal blocks for this partition, there is some choice with the ordering prescription for the matrices. We have found that the following procedure gives sensible results, in particular avoiding edge effects that can introduce extraneous eigenvalues into $Z_\text{cl}$. For the case of the more symmetric circular and vertical cuts, the procedure we describe reduces to the definitions used in the main text.

Firstly, define the $N\times N$ matrix $Z_\text{cl}$ as a projection of a larger fuzzy disk geometry represented by a $2N \times 2N$ matrix $\widetilde{Z}_{cl}$ (the classical ground state of the model with $U(2N)$ symmetry):
\begin{equation}
    Z_\text{cl} = \Theta_{D}\widetilde{Z}_\text{cl}\Theta_{D} \,.
\end{equation}
Here the projection matrix $\Theta_D$ is the same as the circular projector $\Theta_I$ defined in the main text, now projecting to the interior of a circle described by an $N\times N$ matrix.

Secondly, define the projection to the interior of the entanglement cut
\begin{equation}
    \Theta_{\Sigma} \equiv \theta(\widetilde{X}_{cl}^6 + \widetilde{Y}_{cl}^6 - c^6) \,.
\end{equation}
As in the main text, the matrix step function $\theta$ will be easiest to work with in a basis where its argument is diagonal. The off-diagonal blocks are then extracted by taking e.g.
\begin{equation}
    Z_{\Sigma\overline{\Sigma}} = \Theta_{\Sigma}Z_\text{cl}(1-\Theta_{\Sigma}) = \Theta_{\Sigma}\Theta_{D}\widetilde{Z}_\text{cl}\Theta_{D}(1-\Theta_{\Sigma}) \,.
\end{equation}
As the radius of the cut $c$ increases, the cut goes through a few different regimes. It touches the boundary of the disk (the solid gray line in Fig. \ref{fig:scAll}), intersects the disk (e.g. the dashed black line), and then leaves the disk entirely (the solid black line). The remainder of this appendix is devoted to numerical results regarding this off-diagonal block.

\begin{figure}[h!]
    \centering
    \includegraphics[width=0.7\textwidth]{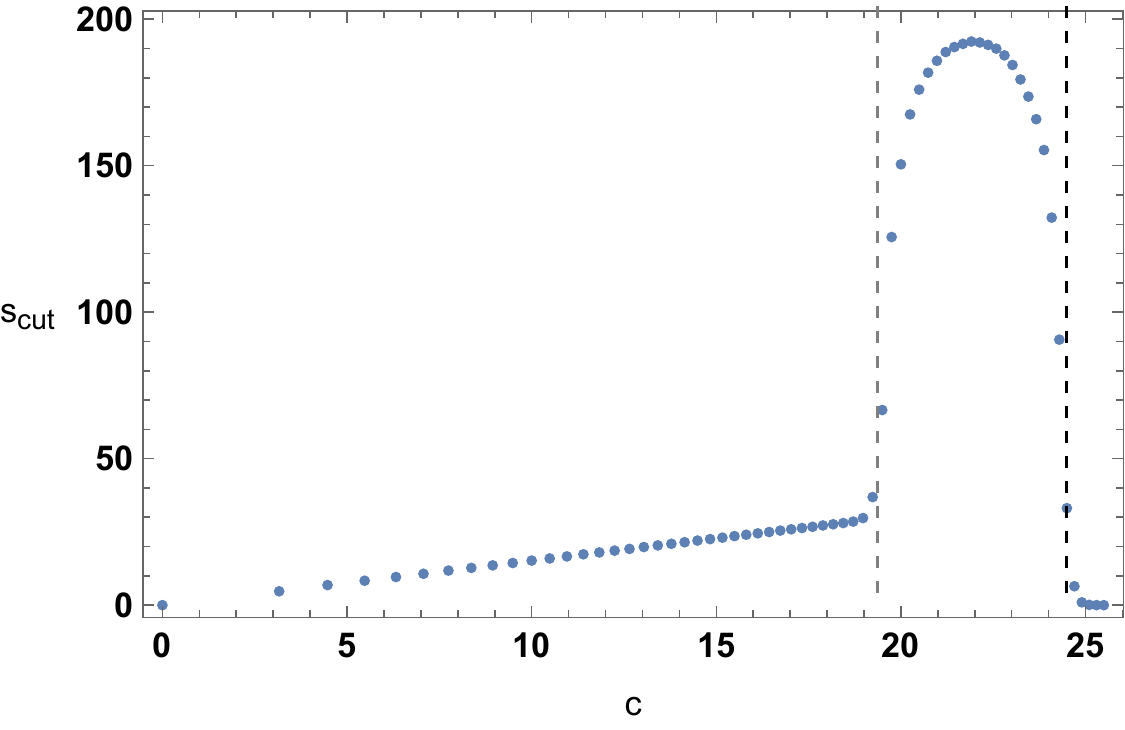}
    \caption{The behavior of the entanglement entropy, $s_\text{cut}$, (unnormalized) as the lengthscale of the square cut increases for $N=600$. The dashed gray vertical line in this figure corresponds to the transition at the solid grey line in Fig. \ref{fig:scAll}, and the dashed black vertical line corresponds to the solid black line where the cut leaves the disc.}
    \label{fig:scEE}
\end{figure}

The behavior of the entanglement entropy is shown in Fig. \ref{fig:scEE}, with the gray line corresponding to the transition where the cut begins to cross the disk boundary, and the black line corresponding to the point where the cut leaves the disk entirely. One can see a linear increase in the entanglement with the lengthscale $c$, corresponding to an area law, followed by a sharp jump to an area law modified by a logarithm, followed by a drop to 0 once the cut leaves the droplet.
To confirm that the behavior after the jump indeed matches that 
of an area law modified by a log, one can keep the location of the cut fixed and vary $N$. This reveals that the entanglement grows as $A \sqrt{N}\log{N}$ for some prefactor $A$. Thus the logarithmic violation of the area law indeed appears precisely when the cut intersects the boundary of the geometry.

\begin{figure}[h!]
    \centering
    \includegraphics[width=0.7\textwidth]{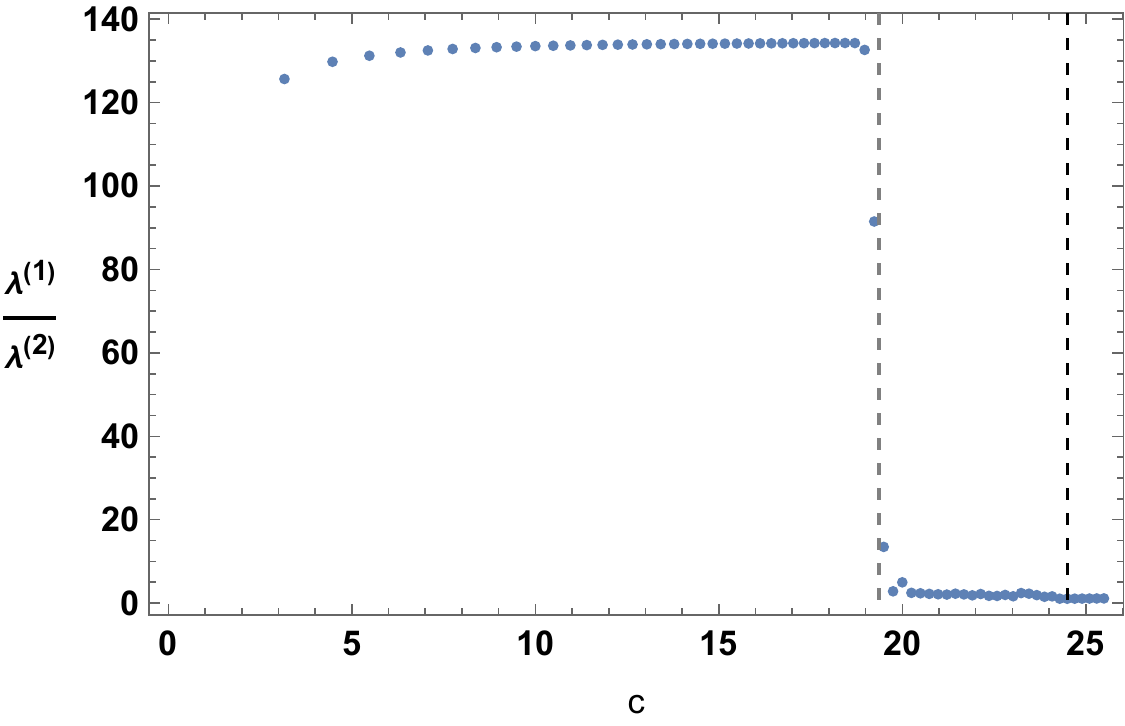}
    \caption{The ratio of the largest to second largest ($\lambda^{(1)}/\lambda^{(2)}$) singular values of the off-diagonal block for the square cut at $N=600$. Before the cut begins crossing the disk boundary there is a clear single singular value that dominates, after the crossing there is a sharp transition to a regime where multiple singular values are relevant.}
    \label{fig:scEvals}
\end{figure}

We may now furthermore verify that the logarithmic violation of the area law occurs simultaneously with the appearance of multiple singular values. Fig. \ref{fig:scEvals} shows the ratio of the largest to second largest singular values of the off-diagonal blocks relevant to the entanglement entropy. While the cut is fully contained inside the disk there is one dominant eigenvalue, once it crosses the boundary there are multiple relevant eigenvalues.

\providecommand{\href}[2]{#2}\begingroup\raggedright\endgroup


\begin{thebibliography}{10}

\bibitem{Ryu:2006bv}
S.~Ryu and T.~Takayanagi, {{Holographic derivation of entanglement entropy from
  AdS/CFT}}, \href{http://dx.doi.org/10.1103/PhysRevLett.96.181602}{Phys. Rev.
  Lett. {\bf 96}, 181602, 2006},
  [\href{http://arxiv.org/abs/arXiv:hep-th/0603001}{{arXiv:hep-th/0603001}}].

\bibitem{Hubeny:2007xt}
V.~E. Hubeny, M.~Rangamani and T.~Takayanagi, {{A Covariant holographic
  entanglement entropy proposal}},
  \href{http://dx.doi.org/10.1088/1126-6708/2007/07/062}{JHEP {\bf 07}, 062,
  2007}, [\href{http://arxiv.org/abs/arXiv:0705.0016}{{arXiv:0705.0016
  [hep-th]}}].

\bibitem{Faulkner:2013ana}
T.~Faulkner, A.~Lewkowycz and J.~Maldacena, {{Quantum corrections to
  holographic entanglement entropy}},
  \href{http://dx.doi.org/10.1007/JHEP11(2013)074}{JHEP {\bf 11}, 074, 2013},
  [\href{http://arxiv.org/abs/arXiv:1307.2892}{{arXiv:1307.2892 [hep-th]}}].

\bibitem{Engelhardt:2014gca}
N.~Engelhardt and A.~C. Wall, {{Quantum Extremal Surfaces: Holographic
  Entanglement Entropy beyond the Classical Regime}},
  \href{http://dx.doi.org/10.1007/JHEP01(2015)073}{JHEP {\bf 01}, 073, 2015},
  [\href{http://arxiv.org/abs/arXiv:1408.3203}{{arXiv:1408.3203 [hep-th]}}].

\bibitem{Polychronakos:2001mi}
A.~P. Polychronakos, {{Quantum Hall states as matrix Chern-Simons theory}},
  \href{http://dx.doi.org/10.1088/1126-6708/2001/04/011}{JHEP {\bf 04}, 011,
  2001},
  [\href{http://arxiv.org/abs/arXiv:hep-th/0103013}{{arXiv:hep-th/0103013}}].

\bibitem{Hellerman:2001rj}
S.~Hellerman and M.~Van~Raamsdonk, {{Quantum Hall physics equals noncommutative
  field theory}}, \href{http://dx.doi.org/10.1088/1126-6708/2001/10/039}{JHEP
  {\bf 10}, 039, 2001},
  [\href{http://arxiv.org/abs/arXiv:hep-th/0103179}{{arXiv:hep-th/0103179}}].

\bibitem{Klebanov:1991qa}
I.~R. Klebanov, {{String theory in two-dimensions}},  in \emph{{Spring School
  on String Theory and Quantum Gravity (to be followed by Workshop)}}, 1991.
\newblock
  [\href{http://arxiv.org/abs/arXiv:hep-th/9108019}{{arXiv:hep-th/9108019}}].

\bibitem{Das:1995vj}
S.~R. Das, {{Geometric entropy of nonrelativistic fermions and two-dimensional
  strings}}, \href{http://dx.doi.org/10.1103/PhysRevD.51.6901}{Phys. Rev. D
  {\bf 51}, 6901--6908, 1995},
  [\href{http://arxiv.org/abs/arXiv:hep-th/9501090}{{arXiv:hep-th/9501090}}].

\bibitem{Das:1995jw}
S.~R. Das, {{Degrees of freedom in two-dimensional string theory}},
  \href{http://dx.doi.org/10.1016/0920-5632(95)00640-0}{Nucl. Phys. B Proc.
  Suppl. {\bf 45BC}, 224--233, 1996},
  [\href{http://arxiv.org/abs/arXiv:hep-th/9511214}{{arXiv:hep-th/9511214}}].

\bibitem{Hartnoll:2015fca}
S.~A. Hartnoll and E.~Mazenc, {{Entanglement entropy in two dimensional string
  theory}}, \href{http://dx.doi.org/10.1103/PhysRevLett.115.121602}{Phys. Rev.
  Lett. {\bf 115}, 121602, 2015},
  [\href{http://arxiv.org/abs/arXiv:1504.07985}{{arXiv:1504.07985 [hep-th]}}].

\bibitem{Das:2020jhy}
S.~R. Das, A.~Kaushal, G.~Mandal and S.~P. Trivedi, {{Bulk Entanglement Entropy
  and Matrices}}, \href{http://dx.doi.org/10.1088/1751-8121/abafe4}{J. Phys. A
  {\bf 53}, 444002, 2020},
  [\href{http://arxiv.org/abs/arXiv:2004.00613}{{arXiv:2004.00613 [hep-th]}}].

\bibitem{Das:2020xoa}
S.~R. Das, A.~Kaushal, S.~Liu, G.~Mandal and S.~P. Trivedi, {{Gauge Invariant
  Target Space Entanglement in D-Brane Holography}},  2020,
  [\href{http://arxiv.org/abs/arXiv:2011.13857}{{arXiv:2011.13857 [hep-th]}}].

\bibitem{Hampapura:2020hfg}
H.~R. Hampapura, J.~Harper and A.~Lawrence, {{Target space entanglement in
  Matrix Models}},  2020,
  [\href{http://arxiv.org/abs/arXiv:2012.15683}{{arXiv:2012.15683 [hep-th]}}].

\bibitem{Banks:1996vh}
T.~Banks, W.~Fischler, S.~H. Shenker and L.~Susskind, {{M theory as a matrix
  model: A Conjecture}},
  \href{http://dx.doi.org/10.1103/PhysRevD.55.5112}{Phys. Rev. D {\bf 55},
  5112--5128, 1997},
  [\href{http://arxiv.org/abs/arXiv:hep-th/9610043}{{arXiv:hep-th/9610043}}].

\bibitem{Berenstein:2002jq}
D.~E. Berenstein, J.~M. Maldacena and H.~S. Nastase, {{Strings in flat space
  and pp waves from N=4 superYang-Mills}},
  \href{http://dx.doi.org/10.1088/1126-6708/2002/04/013}{JHEP {\bf 04}, 013,
  2002},
  [\href{http://arxiv.org/abs/arXiv:hep-th/0202021}{{arXiv:hep-th/0202021}}].

\bibitem{Donnelly:2011hn}
W.~Donnelly, {{Decomposition of entanglement entropy in lattice gauge theory}},
  \href{http://dx.doi.org/10.1103/PhysRevD.85.085004}{Phys. Rev. D {\bf 85},
  085004, 2012}, [\href{http://arxiv.org/abs/arXiv:1109.0036}{{arXiv:1109.0036
  [hep-th]}}].

\bibitem{Buividovich:2008gq}
P.~V. Buividovich and M.~I. Polikarpov, {{Entanglement entropy in gauge
  theories and the holographic principle for electric strings}},
  \href{http://dx.doi.org/10.1016/j.physletb.2008.10.032}{Phys. Lett. B {\bf
  670}, 141--145, 2008},
  [\href{http://arxiv.org/abs/arXiv:0806.3376}{{arXiv:0806.3376 [hep-th]}}].

\bibitem{Casini:2013rba}
H.~Casini, M.~Huerta and J.~A. Rosabal, {{Remarks on entanglement entropy for
  gauge fields}}, \href{http://dx.doi.org/10.1103/PhysRevD.89.085012}{Phys.
  Rev. D {\bf 89}, 085012, 2014},
  [\href{http://arxiv.org/abs/arXiv:1312.1183}{{arXiv:1312.1183 [hep-th]}}].

\bibitem{Ghosh:2015iwa}
S.~Ghosh, R.~M. Soni and S.~P. Trivedi, {{On The Entanglement Entropy For Gauge
  Theories}}, \href{http://dx.doi.org/10.1007/JHEP09(2015)069}{JHEP {\bf 09},
  069, 2015}, [\href{http://arxiv.org/abs/arXiv:1501.02593}{{arXiv:1501.02593
  [hep-th]}}].

\bibitem{Donnelly:2016auv}
W.~Donnelly and L.~Freidel, {{Local subsystems in gauge theory and gravity}},
  \href{http://dx.doi.org/10.1007/JHEP09(2016)102}{JHEP {\bf 09}, 102, 2016},
  [\href{http://arxiv.org/abs/arXiv:1601.04744}{{arXiv:1601.04744 [hep-th]}}].

\bibitem{Tong:2015xaa}
D.~Tong and C.~Turner, {{Quantum Hall effect in supersymmetric Chern-Simons
  theories}}, \href{http://dx.doi.org/10.1103/PhysRevB.92.235125}{Phys. Rev. B
  {\bf 92}, 235125, 2015},
  [\href{http://arxiv.org/abs/arXiv:1508.00580}{{arXiv:1508.00580 [hep-th]}}].

\bibitem{Belin:2019mlt}
A.~Belin, N.~Iqbal and J.~Kruthoff, {{Bulk entanglement entropy for photons and
  gravitons in AdS$_3$}},
  \href{http://dx.doi.org/10.21468/SciPostPhys.8.5.075}{SciPost Phys. {\bf 8},
  075, 2020}, [\href{http://arxiv.org/abs/arXiv:1912.00024}{{arXiv:1912.00024
  [hep-th]}}].

\bibitem{PhysRevLett.96.110404}
A.~Kitaev and J.~Preskill, {Topological entanglement entropy},
  \href{http://dx.doi.org/10.1103/PhysRevLett.96.110404}{Phys. Rev. Lett. {\bf
  96}, 110404, 2006}.

\bibitem{Das:2015oha}
D.~Das and S.~Datta, {{Universal features of left-right entanglement entropy}},
  \href{http://dx.doi.org/10.1103/PhysRevLett.115.131602}{Phys. Rev. Lett. {\bf
  115}, 131602, 2015},
  [\href{http://arxiv.org/abs/arXiv:1504.02475}{{arXiv:1504.02475 [hep-th]}}].

\bibitem{Wong:2017pdm}
G.~Wong, {{A note on entanglement edge modes in Chern Simons theory}},
  \href{http://dx.doi.org/10.1007/JHEP08(2018)020}{JHEP {\bf 08}, 020, 2018},
  [\href{http://arxiv.org/abs/arXiv:1706.04666}{{arXiv:1706.04666 [hep-th]}}].

\bibitem{PhysRevLett.96.110405}
M.~Levin and X.-G. Wen, {Detecting topological order in a ground state wave
  function}, \href{http://dx.doi.org/10.1103/PhysRevLett.96.110405}{Phys. Rev.
  Lett. {\bf 96}, 110405, 2006}.

\bibitem{PhysRevLett.50.1395}
R.~B. Laughlin, {Anomalous quantum hall effect: An incompressible quantum fluid
  with fractionally charged excitations},
  \href{http://dx.doi.org/10.1103/PhysRevLett.50.1395}{Phys. Rev. Lett. {\bf
  50}, 1395--1398, 1983}.

\bibitem{Susskind:2001fb}
L.~Susskind, {{The Quantum Hall fluid and noncommutative Chern-Simons theory}},
   2001,
  [\href{http://arxiv.org/abs/arXiv:hep-th/0101029}{{arXiv:hep-th/0101029}}].

\bibitem{Karabali:2001xq}
D.~Karabali and B.~Sakita, {{Chern-Simons matrix model: Coherent states and
  relation to Laughlin wavefunctions}},
  \href{http://dx.doi.org/10.1103/PhysRevB.64.245316}{Phys. Rev. B {\bf 64},
  245316, 2001},
  [\href{http://arxiv.org/abs/arXiv:hep-th/0106016}{{arXiv:hep-th/0106016}}].

\bibitem{JEVICKI1980511}
A.~Jevicki and B.~Sakita, {The quantum collective field method and its
  application to the planar limit},
  \href{http://dx.doi.org/https://doi.org/10.1016/0550-3213(80)90046-2}{Nuclear
  Physics B {\bf 165}, 511--527, 1980}.

\bibitem{ANDRIC1983307}
I.~Andrić, A.~Jevicki and H.~Levine, {On the large-n limit in symplectic
  matrix models},
  \href{http://dx.doi.org/https://doi.org/10.1016/0550-3213(83)90218-3}{Nuclear
  Physics B {\bf 215}, 307--315, 1983}.

\bibitem{Brezin:1977sv}
E.~Brezin, C.~Itzykson, G.~Parisi and J.~B. Zuber, {{Planar Diagrams}},
  \href{http://dx.doi.org/10.1007/BF01614153}{Commun. Math. Phys. {\bf 59}, 35,
  1978}.

\bibitem{JACKIW1981133}
R.~Jackiw and A.~Strominger, {Wave function(al)s in the large-n limit},
  \href{http://dx.doi.org/https://doi.org/10.1016/0370-2693(81)90966-7}{Physics
  Letters B {\bf 99}, 133 -- 140, 1981}.

\bibitem{SONNENSCHEIN1988752}
J.~Sonnenschein, {Chiral bosons},
  \href{http://dx.doi.org/https://doi.org/10.1016/0550-3213(88)90339-2}{Nuclear
  Physics B {\bf 309}, 752 -- 770, 1988}.

\bibitem{Mazenc:2019ety}
E.~A. Mazenc and D.~Ranard, {{Target Space Entanglement Entropy}},  2019,
  [\href{http://arxiv.org/abs/arXiv:1910.07449}{{arXiv:1910.07449 [hep-th]}}].

\bibitem{Sugishita:2021vih}
S.~Sugishita, {{Target space entanglement in quantum mechanics of fermions and
  matrices}}, \href{http://dx.doi.org/10.1007/JHEP08(2021)046}{JHEP {\bf 08},
  046, 2021}, [\href{http://arxiv.org/abs/arXiv:2105.13726}{{arXiv:2105.13726
  [hep-th]}}].

\bibitem{Holzhey:1994we}
C.~Holzhey, F.~Larsen and F.~Wilczek, {{Geometric and renormalized entropy in
  conformal field theory}},
  \href{http://dx.doi.org/10.1016/0550-3213(94)90402-2}{Nucl. Phys. B {\bf
  424}, 443--467, 1994},
  [\href{http://arxiv.org/abs/arXiv:hep-th/9403108}{{arXiv:hep-th/9403108}}].

\bibitem{Lunin:2000yv}
O.~Lunin and S.~D. Mathur, {{Correlation functions for M**N / S(N) orbifolds}},
  \href{http://dx.doi.org/10.1007/s002200100431}{Commun. Math. Phys. {\bf 219},
  399--442, 2001},
  [\href{http://arxiv.org/abs/arXiv:hep-th/0006196}{{arXiv:hep-th/0006196}}].

\bibitem{Berenstein:2004kk}
D.~Berenstein, {{A Toy model for the AdS / CFT correspondence}},
  \href{http://dx.doi.org/10.1088/1126-6708/2004/07/018}{JHEP {\bf 07}, 018,
  2004},
  [\href{http://arxiv.org/abs/arXiv:hep-th/0403110}{{arXiv:hep-th/0403110}}].

\bibitem{Itzhaki:2004te}
N.~Itzhaki and J.~McGreevy, {{The Large N harmonic oscillator as a string
  theory}}, \href{http://dx.doi.org/10.1103/PhysRevD.71.025003}{Phys. Rev. D
  {\bf 71}, 025003, 2005},
  [\href{http://arxiv.org/abs/arXiv:hep-th/0408180}{{arXiv:hep-th/0408180}}].

\bibitem{Hellerman:2021fla}
S.~Hellerman, D.~Orlando and M.~Watanabe, {{Quantum Information Theory of the
  Gravitational Anomaly}},  2021,
  [\href{http://arxiv.org/abs/arXiv:2101.03320}{{arXiv:2101.03320 [hep-th]}}].

\bibitem{Balasubramanian:2018por}
V.~Balasubramanian, M.~DeCross, J.~Fliss, A.~Kar, R.~G. Leigh and O.~Parrikar,
  {{Entanglement Entropy and the Colored Jones Polynomial}},
  \href{http://dx.doi.org/10.1007/JHEP05(2018)038}{JHEP {\bf 05}, 038, 2018},
  [\href{http://arxiv.org/abs/arXiv:1801.01131}{{arXiv:1801.01131 [hep-th]}}].

\bibitem{Strominger:1996sh}
A.~Strominger and C.~Vafa, {{Microscopic origin of the Bekenstein-Hawking
  entropy}}, \href{http://dx.doi.org/10.1016/0370-2693(96)00345-0}{Phys. Lett.
  B {\bf 379}, 99--104, 1996},
  [\href{http://arxiv.org/abs/arXiv:hep-th/9601029}{{arXiv:hep-th/9601029}}].

\bibitem{Jevicki:1993qn}
A.~Jevicki, {{Development in 2-d string theory}},  in \emph{{Workshop on String
  Theory, Gauge Theory and Quantum Gravity}}, 1993.
\newblock
  [\href{http://arxiv.org/abs/arXiv:hep-th/9309115}{{arXiv:hep-th/9309115}}].

\bibitem{Han:2019wue}
X.~Han and S.~A. Hartnoll, {{Deep Quantum Geometry of Matrices}},
  \href{http://dx.doi.org/10.1103/PhysRevX.10.011069}{Phys. Rev. X {\bf 10},
  011069, 2020},
  [\href{http://arxiv.org/abs/arXiv:1906.08781}{{arXiv:1906.08781 [hep-th]}}].

\bibitem{FRADKIN2002483}
E.~Fradkin, V.~Jejjala and R.~G. Leigh, {Non-commutative chern–simons for the
  quantum hall system and duality},
  \href{http://dx.doi.org/https://doi.org/10.1016/S0550-3213(02)00616-8}{Nuclear
  Physics B {\bf 642}, 483--500, 2002}.

\bibitem{Dorey:2016mxm}
N.~Dorey, D.~Tong and C.~Turner, {{Matrix model for non-Abelian quantum Hall
  states}}, \href{http://dx.doi.org/10.1103/PhysRevB.94.085114}{Phys. Rev. B
  {\bf 94}, 085114, 2016},
  [\href{http://arxiv.org/abs/arXiv:1603.09688}{{arXiv:1603.09688
  [cond-mat.str-el]}}].

\bibitem{Dorey:2016hoj}
N.~Dorey, D.~Tong and C.~Turner, {{A Matrix Model for WZW}},
  \href{http://dx.doi.org/10.1007/JHEP08(2016)007}{JHEP {\bf 08}, 007, 2016},
  [\href{http://arxiv.org/abs/arXiv:1604.05711}{{arXiv:1604.05711 [hep-th]}}].

\bibitem{Minwalla:1999px}
S.~Minwalla, M.~Van~Raamsdonk and N.~Seiberg, {{Noncommutative perturbative
  dynamics}}, \href{http://dx.doi.org/10.1088/1126-6708/2000/02/020}{JHEP {\bf
  02}, 020, 2000},
  [\href{http://arxiv.org/abs/arXiv:hep-th/9912072}{{arXiv:hep-th/9912072}}].

\bibitem{Andric_1988}
I.~Andric and V.~Bardek, {1/n corrections in calogero-type models using the
  collective-field method},
  \href{http://dx.doi.org/10.1088/0305-4470/21/13/009}{Journal of Physics A:
  Mathematical and General {\bf 21}, 2847--2853, 1988}.

\end{thebibliography}
\end{document}